\documentclass[preprint2]{aastex}
\usepackage{graphicx}
 
\newcommand{\km}{\,\mbox{km}\,\mbox{s}^{-1}}

\def\i{\,{\small I}}
 
\def\iii{\,{\small III}}

\shorttitle{Nuclei and inner polar disks} \shortauthors{Sil'chenko}

\begin{document}

\title{Stellar nuclei and inner polar disks in lenticular galaxies}

\author{Olga K. Sil'chenko} \affil{Sternberg Astronomical Institute, M.V.
Lomonosov Moscow State University, Moscow, 119992 Russia, and Isaac Newton
Institute, Chile, Moscow Branch} \email{olga@sai.msu.su}

\begin{abstract} 

I analyze statistics of the stellar population properties for stellar nuclei and bulges of nearby 
lenticular galaxies in different environments by using panoramic spectral data
of the integral-field spectrograph SAURON retrieved from the open archive of Isaac Newton Group.
I estimate also the fraction of nearby lenticular galaxies having inner polar
gaseous disks by exploring the volume-limited sample of early-type galaxies of the
ATLAS-3D survey. By inspecting the two-dimensional velocity fields of the stellar and gaseous
components with running tilted-ring technique, I have found 7 new cases of
the inner polar disks. Together with those, the frequency of inner polar disks in
nearby S0 galaxies reaches 10\%\ that is much higher than the frequency of large-scale
polar rings. Interestingly, the properties of the nuclear stellar populations in the
inner polar ring hosts are statistically the same as those in the whole S0 sample
implying similar histories of multiple gas accretion events from various directions.

\end{abstract}

\keywords{ galaxies: elliptical and lenticular --- galaxies: ISM --- galaxies:
kinematics and dynamics --- galaxies: evolution}

\section{Introduction}

The outer gas accretion is now recognized as a main driver of disk galaxy evolution:
neither prolonged star formation observed in disks of spiral galaxies, nor observed chemical 
abundances in their stars can be explained without continuous gas supply from outside. 
Indeed, the gas depletion time in nearby spirals is found to concentrate tightly around the value of only
2--3~Gyr \citep{bigiel} while the solar abundance ratios of the disk stellar populations imply
the duration of the continuous star formation of more than 3~Gyr. The `G-dwarf paradox' and
the absence of the age-metallicity correlation in the thin stellar disk of our own Galaxy \citep{tosi,chiosi}
as well as the lowered effective oxygen yield in the disks of other spiral galaxies \citep{pilyugin,dalcanton}
require such accretion. However, direct observational findings of outer-gas accretion signatures are
rather rare despite we expect these events to happen daily. Perhaps, it would be easier to search
for consequences of outer gas accretion in early-type disk galaxies, namely, in S0s, where
own gas of the galaxies, usually absent, does not prevent rather long-lived kinematical
misalignments between the stellar disks and the accreted gaseous subsystems.

One of the most bright phenomena betraying the outer-gas accretion events are {\it inner} polar rings/disks 
of ionized gas which are embedded deeply into the bulge-dominated area. The presence of some minor species
with a decoupled momentum can be explained only by such accretion. The first finding of the inner polar disk was noted 
by  \citet{bettoni} in SB0-galaxy NGC~2217 from the multiple long-slit cross-sections of this barred 
galaxy; the complex gas kinematics was explained by a strong warp of the gas rotation plane 
in the center of the galaxy, such that the central gas rotation proceeded in the plane orthogonal 
to the stellar rotation plane and also to the bar major axis. After the rise of the era of integral-field 
spectroscopy, the inner polar disks were found also in many unbarred galaxies, in particular: 
in Sb-galaxy NGC~2841 \citep{silvb97},
in Sb galaxy NGC~7742 \citep{n7742we}, and in isolated Sa-galaxy NGC~7217 \citep{we7217}. They were 
detected exclusively through the kinematical analysis: the integral-field spectroscopy allowed to determine 
spatial orientations of the rotation axes both for the stellar and ionized-gas components, and if these 
rotation axes appeared to be mutually orthogonal, the presence of the inner polar disk could be claimed. 
Sometimes inner polar rings of the ionized gas could be seen as dust lanes aligned along the minor axes 
of the isophotes -- in such a way we found 8 inner polar disks in lenticular galaxies whose high-resolution
images were provided by the Hubble Space Telescope (HST) \citep{polars0}. The first list of 17 galaxies
which were claimed to possess inner polar disks was presented by \citet{inpoldisk}.

Now a few dozen of the inner polar gaseous disks are already known, and the time for their statistics
has come. \citet{moisrev} has assembled a list of 47 inner polar disks reported by various
authors before 2012, and has presented some general properties of the inner polar disks and their host galaxies.
Firstly, these disks are indeed polar: though all the gaseous rings whose rotation axes are inclined to
the stellar rotation axes by more than 50 deg have been considered, the distribution of the mutual
inclinations peaks strongly at 90 deg. Secondly, they can be met mostly in early-type galaxies:
more than the half of all known inner polar disks belong to (mostly) lenticular and elliptical galaxies; 
however a few ones belonging to very late-type dwarfs are also known. Typical radii of the inner polar
disks range from 0.2 to 2.0 kpc; the outer boundary is quite real betraying the relation of the
inner polar disks to bulge-dominated areas, while the inner limit results from finite
spatial resolution of the ground-based integral-field spectroscopy. 

However, the review by \citet{moisrev} operated with a very heterogeneous sample of casual
observational findings. A particular question remains unanswered: what is a frequency of the inner 
polar disks/rings in the whole ensemble of nearby lenticular galaxies? The answer would help
to specify the geometry of the outer-gas accretion and so to identify its sources. In the 
case of isotropic accretion we would be able to estimate the theoretical fraction of inner polar rings
by taking into account their dynamical evolution -- precession and sinking to the main symmetry planes. 
If the theoretical estimates diverge with the observational statistics, it may be a hint to anisotropic
accretion-source distribution -- e.g. accretion from a single neighbor galaxy or multiple minor mergers from the 
satellite plane. I have undertaken a further attempt to increase the number of the known inner polar disks by using
the possibility provided by the integral-field spectral data for a sample of early-type
galaxies which have been obtained with the IFU SAURON \citep{sauron} in the frame of the ATLAS-3D survey
\citep{atlas3d_1}. The ATLAS-3D sample was volume-limited and complete above the absolute magnitude of $M_K=-21.5$, 
so I hoped to estimate reliably the frequency of inner polar disks in nearby lenticular galaxies from these data.
The data of the ATLAS-3D survey are free for retrieving from the Isaac Newton Group (ING)
Archive (CASU Astronomical Data Centre at the Institute of Astronomy, Cambridge) after the end of the
proprietary period, and I have retrieved the raw data for about 150 lenticular galaxies, to analyze
the kinematics of the stellar and ionized-gas components in the central parts of these galaxies
and to search for new cases of the inner polar disks. 

\section{The data of the ATLAS-3D survey}

The integral-field spectrograph SAURON was operating at the 4.2m William Herschel Telescope
belonging to the Isaac Newton Group (ING) of telescopes on La Palma. It worked in so called `TIGER-mode' 
giving about 1500 spectra simultaneously, each for a $0.94^{\prime \prime} \times 0.94^{\prime \prime}$ 
square element (`spaxel') from a (central) part of a galaxy. A total set of spectra covers the area of 
$41^{\prime \prime} \times 33^{\prime \prime}$. The spectral range of the unit is rather narrow, 
4800--5350~\AA, and its spectral resolution is fixed since 2007 at about 4.3~\AA.

There were two surveys of nearby early-type galaxies with the SAURON. The first one which started
in 1999 and finished in 2004, involved 72 galaxies, among those 48 early-type ones and 24 spirals \citep{sauron2}, 
The second one undertaken in 2007--2008 has added more early-type galaxies, including dwarfs, 
to complete the volume-limited ($D<42$ Mpc) sample \citep{atlas3d_1}. The total sample of early-type 
galaxies investigated in these two surveys includes 260 objects, and
200 of them are lenticular galaxies. For my analysis I have selected a subsample of 143 S0
galaxies which have been observed in 2007--2008, because the design of the SAURON has been
successfully modified just before 2007. Due to a new volume-phase grating, 
the last data show much less shift of the spectral range over the field of view, 
and the problems of the first SAURON survey with the absorption lines H$\beta$ and $\mbox{Fe}$\i$\lambda$5270 
which are at the edges of the spectral range exposed \citep{sauron2,sauron6} are now overcome.

\section{Analysis of the data}

\begin{figure}
\includegraphics[width=8cm]{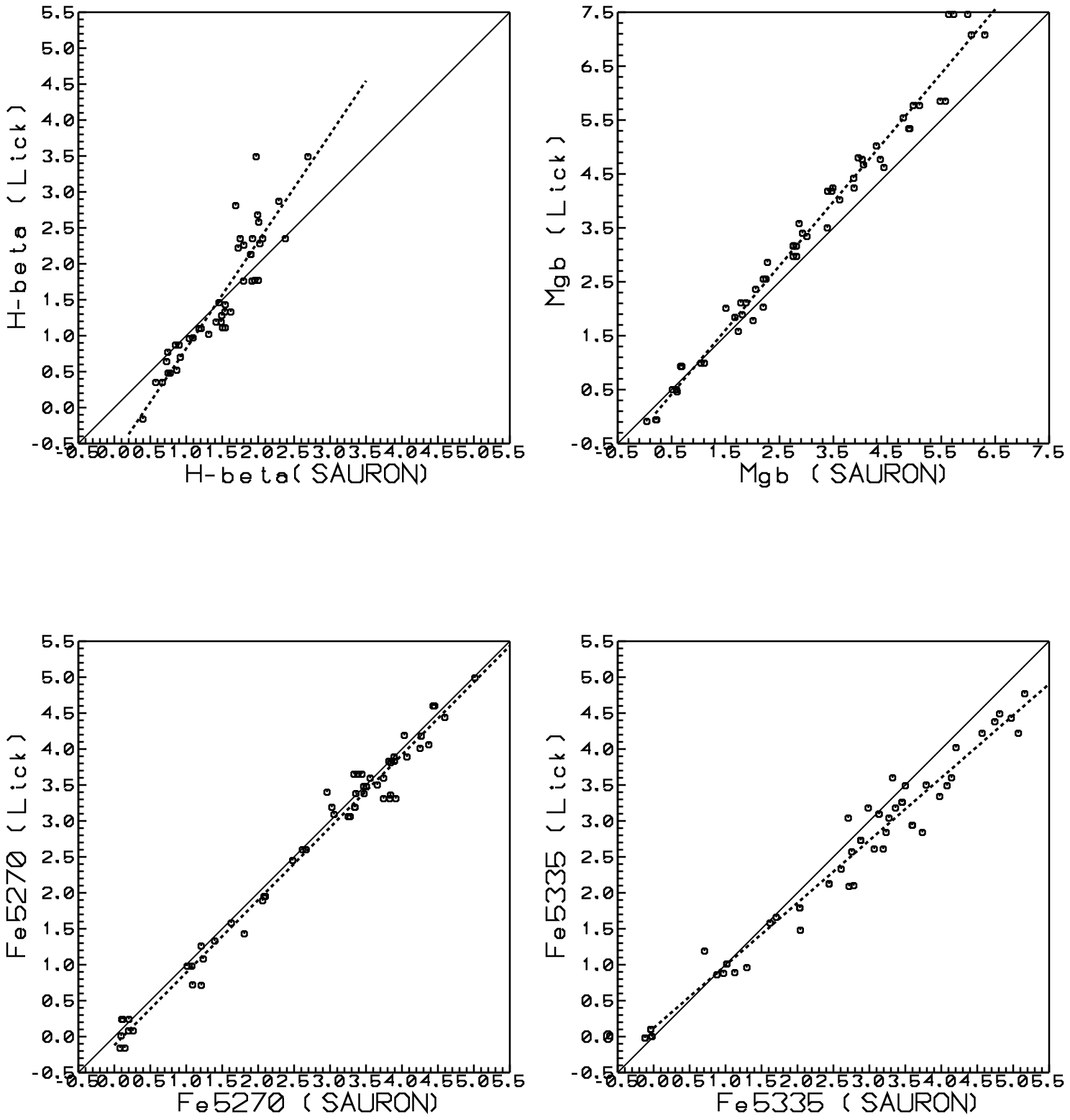}
\caption{One of my calibrations of the Lick index system by using standard stars from the list
of \citet{woretal}: straightforward index measurements from the SAURON spectra are confronted against the standard
Lick values. Signs mark the measurements of individual stars, straight solid lines are equality
lines, and by the dashed lines I show the least-square fit which was used to calibrate the instrumental 
SAURON measurements onto the standard Lick index system. These particular regressions are obtained for the 
observational run of April 19--25, 2007. 
}
\label{indsystem}
\end{figure}

The raw data, both scientific frames and calibration frames, have been retrieved for these
143 S0 galaxies from the ING Archive and reduced in our manner which is described in my paper by 
\citet{sil5sau}. The spectra of a few dozen stars, mostly F--K giants, from the Lick standard list
\citep{woretal} have been also reduced and used, firstly, for cross-correlation with the
galactic spectra, to derive stellar line-of-sight (LOS) velocity fields and stellar velocity dispersion
fields, and secondly, to establish a system of the Lick indices \citep{woretal} used for diagnostics 
of the stellar population properties. The Lick index system was checked for every observational run
separately; an example of such calibration for the second half of the April-2007 observational run
can be seen in Fig.~\ref{indsystem}. The intrinsic scatter of the stars around the calibration lines 
is about 0.2--0.3~\AA, which is consistent with the declared accuracy
of the standard star measurements in \citet{woretal}; our internal accuracy of the index measurements
estimated from the repeatitive observations of the same stars is better than 0.05~\AA. 
As one can see, the spectral range of the observations in 2007
was optimal to measure the indices H$\beta$, Mgb, and Fe5270; and even Fe5335 for the zero redshift
is calibrated appropriately though cannot be used in galaxies with non-zero redshifts. 
After subtracting stellar continuum, I extracted also emission-line 
spectra which were used to calculate LOS velocities of the ionized gas by measuring baricenter positions 
of the strongest emission line in this spectral range, [O\iii]$\lambda$5007~\AA.

These LOS-velocity maps, both for the stellar ($s$) and ionized-gas ($g$) components, have been analyzed 
in the frame of tilted-ring approach in a particular modification by Alexei Moiseev \citep[ the DETKA software]{mois_aa}.
This technique allows to estimate parameters of the spatial orientation of the rotation planes
for both components, namely, the inclinations of the rotation planes $i_{s,g}$ and the position angles of lines 
of nodes, $PA_{s,g}$. After that, the mutual inclinations of the stellar and gaseous rotation
planes have been calculated by using the following formula \citep{moisrev}:
$$
\cos \Delta i = \pm \cos (PA_s - PA_g) \sin i_s \sin i_g + \cos i_s \cos i_g.
$$
The equation has two solutions because typically we do not know which side of a
galactic disk is closer to us. If for the regularly rotating stellar and gaseous components
I obtain both possible mutual inclinations $\Delta i >50$~deg, I consider the circumnuclear 
ionized-gas component to be an inner polar disk. Please note that to determine the mutual
inclination of two rotation planes, one needs to know not only kinematical misalignment,
$(PA_s - PA_g)$, but also inclinations of the planes to the sky plane, $i_s$ and $i_g$.
Only in the case when both planes are seen nearly edge-on, the term with $(PA_s - PA_g)$
is dominant; alternatively, if one of the planes is close to face-on, the kinematical misalignment
does not take any part in the mutual inclination determination. It is just why I can state that
the ATLAS-3D team \citep{atlas3d_2,atlas3d_10} has not established the presence of inner polar disks 
in the galaxies of their
sample: they have measured only kinematical misalignment $(PA_s - PA_g)$ by applying the kinemetry
approach, and only the statistics of this misalignment has been considered \citep{atlas3d_10};
while to determine the intrinsic orientations of the rotation planes, a full tilted-ring
approach is needed. It is made in the present work.

\begin{figure*}
\begin{tabular}{c c}
 \includegraphics[width=8cm]{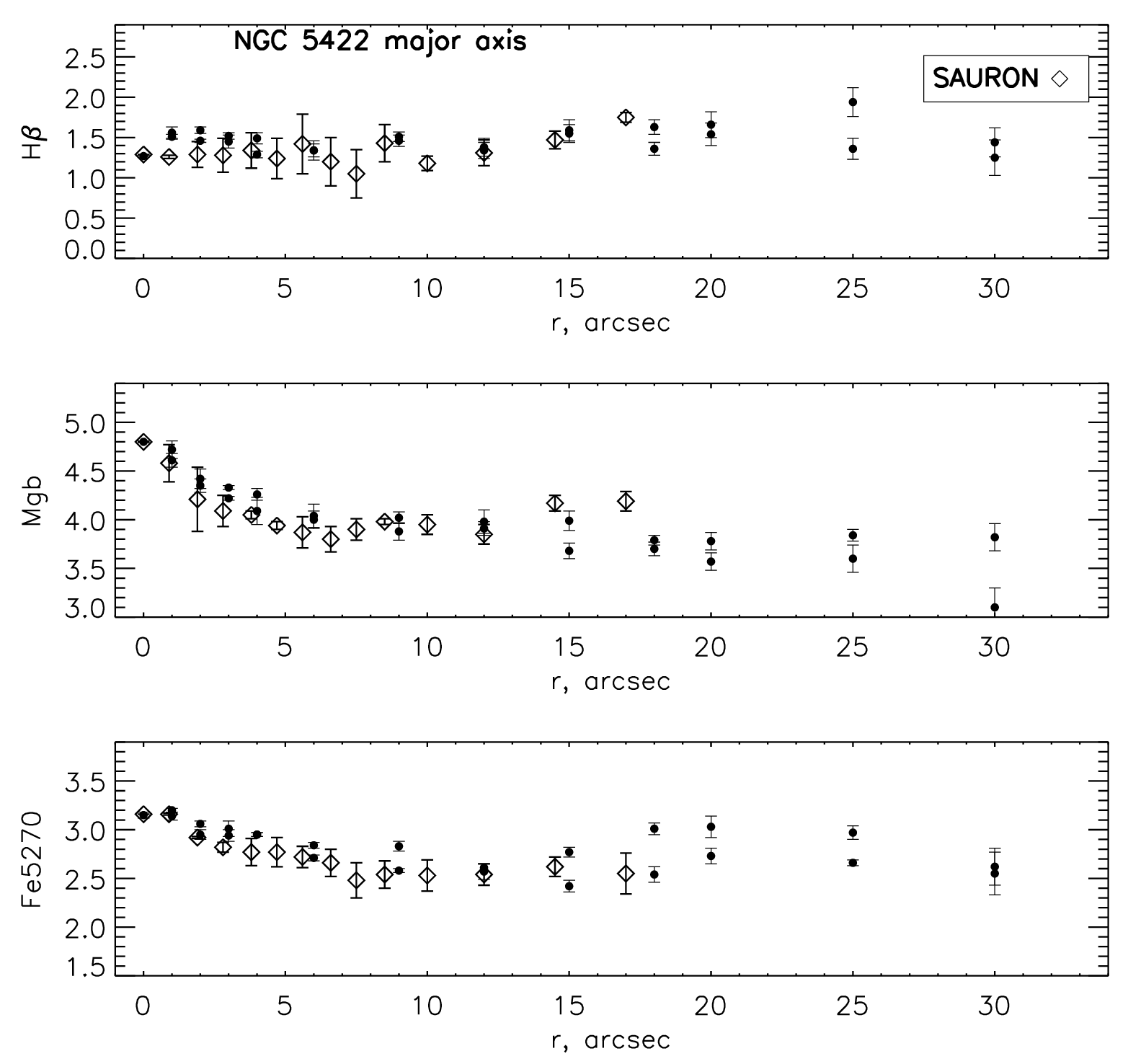} &
 \includegraphics[width=8cm]{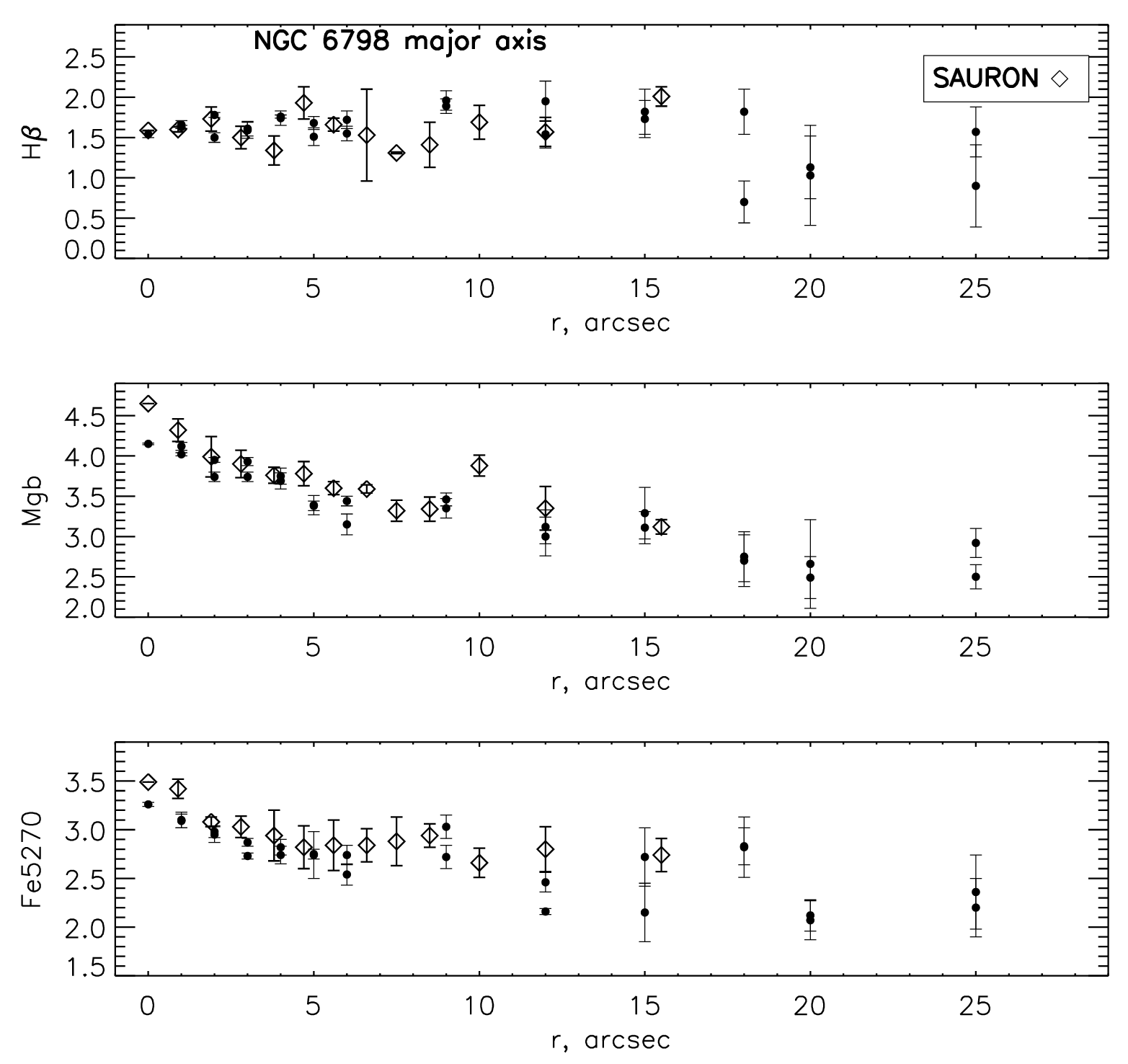} \\
\end{tabular}
\caption{The comparison of the Lick index profiles obtained by using the SAURON data with those
observed by us at the Russian 6m telescope with the long-slit spectrograph SCORPIO. Two halves
of our long-slit cross-sections symmetric with respect to the nuclei are superimposed while
the SAURON data are averaged in the radius bins.
}
\label{vsscorpio}
\end{figure*}

\begin{figure*}
\begin{tabular}{c c}
 \includegraphics[width=8cm]{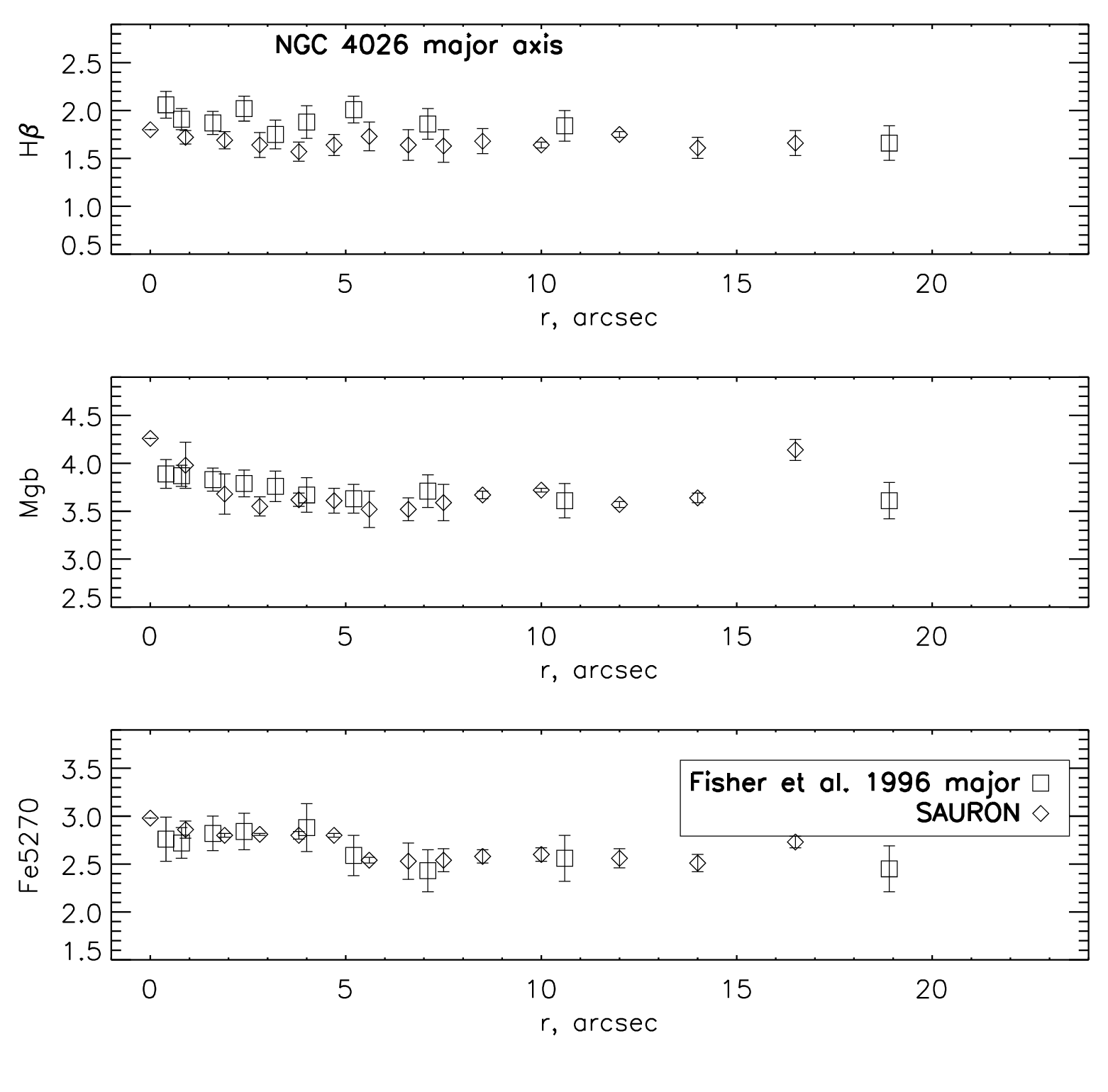} &
 \includegraphics[width=8cm]{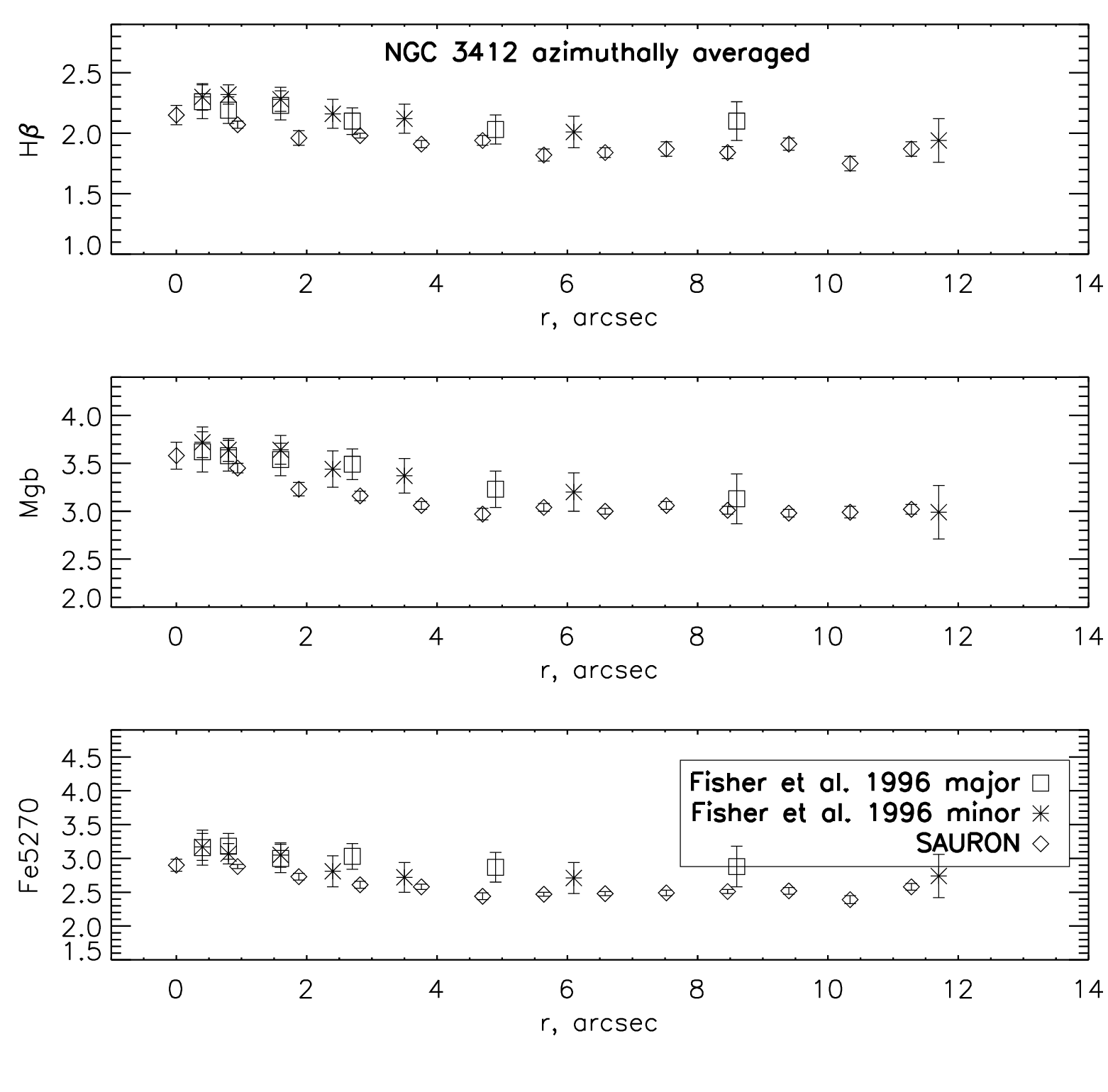} \\
\end{tabular}
\caption{The comparison of the Lick index profiles obtained by using the SAURON data with those
observed along the major and minor axes of the galaxies with a long-slit spectrograph by \citet{fisher}.
}
\label{vsfisher}
\end{figure*}

Also the maps of the Lick indices H$\beta$, Mgb, and Fe5270 calibrated into the standard Lick system
and corrected for the stellar velocity dispersion broadening as described in \citet{sil5sau}, 
have been calculated for all the S0s without too strong emission lines in the centers --
$EW(\mbox{[O\iii]}\lambda 5007) <2$~\AA. If the emission lines were less strong, I have rectified 
the H$\beta$ indices by applying an approach proposed by \citet{trager00}: a correction 
of $\Delta \mbox{H}\beta = 0.6 EW(\mbox{[O\iii]}\lambda 5007)$ was added
to the measured value of the H$\beta$ index to take into account contamination by ionized-hydrogen emission.
As I have found earlier by studying a sample of S0 galaxies with the spectra in the full optical range \citep{s0mpfs},
under this condition, $EW(\mbox{[O\iii]}\lambda 5007) <2$~\AA, the H$\beta$ correction for the emission contamination
is the same being calculated either through the empirical relation with [O\iii] or through the Balmer decrement 
applied to the H$\alpha$ emission-line measurements.
When emission lines are strong -- $EW(\mbox{[O\iii]}\lambda 5007) >2$~\AA, -- the correction through [O\iii] works poorly.
However, for a few galaxies with rather strong emission lines we have our own long-slit spectra
covering the red spectral range -- for NGC~3599 \citep{n3599}, for NGC~7743 \citep{n7743}, and for IC~560 \citep{ic560}. 
It makes possible to identify gas excitation mechanism by using the BPT-diagrams \citep{bpt} and to determine 
the H$\beta$ correction for the emission through the H$\alpha$ emission-line equivalent width estimates 
by applying a model Balmer decrement suitable for the particular excitation mechanism.
If the gas is fully excited by young stars, the model from \citet{burgess} is used,
and if LINER- or shock-like excitation is added, we apply the correction 
$\Delta \mbox{H}\beta \approx 0.25 EW(\mbox{H}\alpha \, \mbox{emis})$ --
the empirical Balmer decrement found by \citet{sts2001} for a large sample of disk galaxies which corresponds
to the mixed gas excitation.

After obtaining the corrected index maps, for the further analysis of the data, radial profiles of the 
Lick indices were derived by averaging indices over the rings with the width of one spaxel for the
galaxies moderately inclined to the line of sight (with the isophote axis ratio less than 1.5), or by superimposing 
a digital slit with a width of $2.5^{\prime \prime} - 3^{\prime \prime}$ (a compromise between the spatial
resolution defined by the seeing and a necessity of sufficient signal-to-noise ratio) to derive
major-axis index profiles for the strongly inclined galaxies.
Since some brighter members of the sample have been already studied as concerning their Lick index distributions,
I can compare the measurements of the Lick indices from the SAURON data with some well-calibrated long-slit 
spectral data. Figure~\ref{vsscorpio} shows the comparison of the major-axis index profiles for two galaxies, 
NGC~5422 and NGC~6798, observed by us earlier with the reducer SCORPIO \citep{scorpio2} of the Russian 6m 
telescope and published elsewhere \citep{isoabul,jenampro}. Figure~\ref{vsfisher} shows the comparison of two 
variants of the Lick index profiles derived from the SAURON maps -- ones averaged over full-azimuth rings 
for NGC~3412 and the others cut by the digital slit along the equator ($PA=178^{\circ}$) for the edge-on galaxy
NGC~4026 -- with the long-slit data by \citet{fisher}. One can see an overall agreement of the profiles; 
so I conclude that I have succeeded to measure the Lick indices in the SAURON data without noticeable systematic shifts.

\section{The characteristics of the nuclear and bulge stellar populations}

In this section I present two separate sets of the data -- the Lick indices for the nuclei, $r<1^{\prime \prime}$, 
and the Lick indices for the bulges taken over one-spaxel rings fixed at $r_e(bul)$ for slightly inclined galaxies, 
$i<50^{\circ}$, or at the distance of $r_e(bul)$ from the nuclei along the minor axes for nearly edge-on galaxies. 
Mean (close to luminosity-weighted) 
parameters of the stellar populations were calculated in the frame of the evolutionary synthesis models of Simple 
Stellar Populations (SSP). The SSP-equivalent ages, metallicities, and magnesium-to-iron ratios were 
determined by confronting Mgb vs Fe5270 and H$\beta$ vs [MgFe52]$\equiv \sqrt{(\mbox{Mgb} \times \mbox{Fe5270})}$; 
I confronted my index measurements to the SSP-models by \citet{thomod} which are calculated for a set of
the magnesium-to-iron ratios. The results -- the Lick indices and the SSP-equivalent ages and metallicities -- are
given in the Tables~1 and 2. The typical internal (statistical) accuracy of the indices in the Tables~1 and 2 is 
around 0.1~\AA; subsequently, the internal accuracy of the ages is 0.5~Gyr for young populations,
$T<3$~Gyr, 1~Gyr for intermediate ages, $3< T <8$~Gyr, and 3~Gyr for older ones. The accuracy
of the metallicities is about 0.1~dex.

\begin{deluxetable}{lcccccc}
\tablenum{1}
\tablewidth{16cm}
\tablewidth{0pt}
\tabletypesize{\small}
\tablecaption{The Lick indices and SSP-equivalent parameters for the nuclei.}
\tablehead{
\colhead{Galaxy\tablenotemark{a}}  &  
\colhead{Eq. width of [OIII]$\lambda$5007,\AA} &
\colhead{H$\beta$}  &
\colhead{Mgb}  &
\colhead{Fe5270} &
\colhead{T, Gyr} &
\colhead{[Z/H], dex}
}
\startdata
 N448    &     0.19   &   1.83  &  4.42 &  3.06  &    4   &     +0.5 \\    
 N502    &     0.12   &   1.96  &  4.80 &  3.40  &    3   &     +0.8  \\   
 N509    &     0.84   &   3.37  &  2.38 &  2.32  &    1   &     +0.4  \\   
 N516    &     0.68   &   2.25  &  2.90 &  2.82  &    2   &     +0.2  \\  
 N525    &     0.24   &   2.18  &  3.38 &  2.53  &    3   &     +0.1 \\   
N1121    &     0      &   1.59  &  4.98 &  2.71  &   14   &      0.0 \\         
N1248    &     0.29   &   2.40  &  3.49 &  2.84  &    1.5 &     +0.5  \\        
N1289    &     0.73   &   1.92  &  3.70 &  2.86  &    2   &     +0.4 \\       
N1665    &     0.30   &   2.31  &  3.84 &  3.18  &    2   &     +0.7 \\
N2481    &     0.41   &   1.55  &  4.39 &  3.04  &    5   &     +0.4  \\      
N2577    &     0.27   &   1.48  &  5.02 &  2.88  &    8   &     +0.3  \\       
N2592    &     0      &   1.49  &  5.11 &  2.82  &   12   &     +0.2  \\      
N2697    &     0.52   &   1.94  &  3.62 &  3.13  &    2   &     +0.5  \\       
N2698    &     0      &   1.84  &  4.93 &  2.77  &    5   &     +0.4  \\       
N2824    &     1.25   &   1.00  &  3.36 &  2.36  &   11   &     -0.2  \\      
N2852    &     0.96   &   1.31  &  5.20 &  2.56  &    5   &     +0.3  \\      
N2859    &     0.31   &   1.59  &  4.73 &  3.24  &    5   &     +0.5  \\     
N2880    &     0.28   &   2.20  &  4.08 &  3.06  &    2   &     +0.7 \\        
N2950    &     0.14   &   2.14  &  4.66 &  3.08  &    2   &     +0.7  \\        
N2962    &     0.80   &   1.37  &  5.23 &  3.24  &    4   &     +0.7  \\       
N3098    &     0.24   &   1.88  &  3.16 &  2.64  &    5   &     -0.1  \\      
N3182    &     3.67   &   1.56  &  4.88 &  3.36  &    3   &     +0.8  \\        
N3230    &     0.10   &   1.42  &  5.04 &  2.92  &   12   &     +0.2  \\       
N3245    &     0.62   &   0.24  &  4.73 &  2.66  &   15   &     +0.0  \\       
N3248    &     0.64   &   2.51  &  3.78 &  3.03  &    1   &     +1.0  \\       
N3301    &     0.44   &   2.10  &  3.45 &  2.48  &    3   &     +0.1  \\      
N3400    &     0.39   &   2.08  &  3.65 &  3.17  &    2   &     +0.7 \\      
N3412    &     0.18   &   2.15  &  3.58 &  2.90  &    2   &     +0.4  \\      
N3457    &     0.64   &   2.18  &  3.24 &  2.43  &    2   &     +0.1  \\      
N3458    &     0      &   1.59  &  4.53 &  2.95  &   11   &     +0.2  \\      
N3499    &     0.32   &   2.18  &  3.61 &  3.10  &    2   &     +0.5  \\      
N3530    &     0.17   &   1.79  &  3.56 &  2.77  &    6   &      0.0  \\       
N3595    &     0.54   &   1.80  &  4.28 &  2.75  &    3   &     +0.4 \\        
N3599    &     2.93   &   1.95  &  3.37 &  2.89  &    2   &     +0.4  \\        
N3607    &     0.34   &   1.24  &  5.01 &  3.03  &   12   &     +0.3  \\       
N3610    &     0.34   &   2.10  &  4.32 &  2.83  &    2   &     +0.7  \\       
N3613    &     0      &   1.70  &  4.82 &  2.99  &    7   &     +0.4  \\      
N3619    &     1.19   &   1.49  &  4.56 &  3.18  &    3   &     +0.7  \\       
N3626    &     0.53   &   2.76  &  3.31 &  2.72  &    1   &     +0.8  \\       
N3630    &     0      &   1.77  &  4.61 &  3.22  &    5   &     +0.5  \\       
N3648    &     0.81   &   1.04  &  5.75 &  3.18  &    7   &     +0.6  \\       
N3658    &     0.10   &   2.03  &  4.65 &  3.47  &    3   &     +0.8  \\        
N3665    &     0.28   &   1.16  &  4.86 &  3.26  &   15   &     +0.3  \\       
N3674    &     0      &   1.57  &  4.94 &  2.97  &   10   &     +0.3  \\       
N3694    &     1.73   &   0.74  &  3.36 &  2.20  &   12   &     -0.3  \\      
N3757    &     0.35   &   1.70  &  3.65 &  2.90  &    5   &     +0.15 \\         
N3796    &     0.38   &   2.91  &  2.68 &  2.22  &    1.5 &     +0.1  \\        
N3838    &     0.17   &   2.02  &  3.90 &  2.68  &    3.5 &     +0.2  \\        
N3941    &     0.50   &   1.90  &  4.81 &  3.36  &    2   &     +0.8  \\        
N3945    &     0.50   &   1.45  &  5.35 &  3.44  &    4   &     +0.8  \\       
N4026    &     0.19   &   1.80  &  4.26 &  2.98  &    4   &     +0.3  \\       
N4036    &     1.42   &   0.16  &  5.42 &  2.87  &   15   &     +0.2  \\      
N4078    &     0.10   &   1.69  &  4.41 &  2.78  &    8   &     +0.15 \\       
N4111    &     0.58   &   2.11  &  4.09 &  2.91  &    2   &     +0.7 \\        
N4124    &     0.48   &   2.84  &  2.67 &  2.23  &    1.5 &     +0.1  \\         
N4143    &     0.62   &   0.85  &  5.60 &  3.04  &   15   &     +0.3 \\         
N4179    &     0.10   &   2.26  &  4.63 &  3.21  &    2   &     +0.8  \\      
N4191    &     0.71   &   1.79  &  4.32 &  3.01  &    2   &     +0.7 \\         
N4215    &     0.18   &   2.26  &  4.07 &  2.75  &    2   &     +0.5  \\        
N4233    &     0.84   &   1.69  &  5.30 &  3.32  &    3   &     +1.0  \\        
N4249    &     0.40   &   2.07  &  4.00 &  3.01  &    2   &     +0.7 \\        
N4251    &     0.20   &   2.03  &  4.15 &  2.99  &    3   &     +0.5  \\       
N4255    &     0.40   &   1.52  &  4.40 &  2.77  &    8   &     +0.15 \\        
N4259    &     0.38   &   1.83  &  4.12 &  2.80  &    4   &     +0.3  \\      
N4264    &     0.26   &   2.15  &  4.10 &  3.20  &    2   &     +0.8  \\       
N4267    &     0      &   1.60  &  5.09 &  2.95  &    9   &     +0.4 \\         
N4281    &     0.08   &   1.81  &  4.92 &  3.15  &    4   &     +0.6 \\        
N4324    &     1.33   &   1.83  &  4.00 &  3.23  &    5   &     +0.4 \\        
N4340    &     0.28   &   2.19  &  4.11 &  3.61  &    2   &     +1.0 \\         
N4342    &     0      &   1.42  &  5.53 &  2.97  &   12   &     +0.4 \\        
N4346    &     0.18   &   1.96  &  4.03 &  3.14  &    3   &     +0.5 \\        
N4350    &     0.13   &   1.37  &  4.66 &  3.19  &   12   &     +0.3 \\       
N4371    &     0.24   &   1.69  &  4.33 &  3.06  &    5   &     +0.4 \\        
N4377    &     0      &   1.86  &  4.31 &  2.77  &    6   &     +0.2 \\        
N4417    &     0.15   &   1.73  &  4.19 &  2.95  &    6   &     +0.2 \\         
N4429    &     0.38   &   1.75  &  4.75 &  3.04  &    3   &     +0.6 \\         
N4434    &     0.17   &   1.79  &  4.77 &  3.14  &    3   &     +0.6 \\        
N4442    &     0.35   &   1.59  &  5.58 &  3.35  &    4   &     +0.8 \\        
N4452    &     0.29   &   1.88  &  3.15 &  2.69  &    5   &      0.0 \\        
N4461    &     0.11   &   1.95  &  5.10 &  3.56  &    3   &     +1.0 \\         
N4474    &     0.30   &   2.13  &  3.65 &  2.61  &    4   &     +0.1 \\        
N4476    &     0.35   &   2.40  &  2.75 &  2.23  &    2   &     -0.05 \\       
N4483    &     0.33   &   2.08  &  4.36 &  2.89  &    2   &     +0.7 \\        
N4503    &     0.15   &   1.92  &  4.99 &  3.57  &    3   &     +1.0 \\         
N4521    &     0.47   &   1.38  &  5.03 &  3.35  &    5   &     +0.5 \\         
N4528    &     0.32   &   2.13  &  3.73 &  2.45  &    3   &     +0.2 \\         
N4578    &     0.20   &   1.87  &  4.94 &  2.95  &    3   &     +0.7 \\         
N4596    &     0.15   &   1.37  &  4.66 &  3.19  &   12   &     +0.3 \\         
N4608    &     0.12   &   1.74  &  4.29 &  2.84  &    7   &     +0.2 \\         
N4612    &     0.44   &   2.77  &  3.41 &  3.01  &    1   &     +1.0 \\         
N4623    &     0.22   &   2.15  &  3.70 &  2.97  &    2   &     +0.5 \\        
N4624    &     0.27   &   1.58  &  4.74 &  3.12  &    5   &     +0.4 \\         
N4638    &     0.20   &   2.15  &  4.45 &  3.29  &    2   &     +1.0 \\         
N4643    &     0.36   &   1.97  &  4.50 &  3.02  &    3   &     +0.7 \\         
N4690    &     0.40   &   1.78  &  2.55 &  3.49  &    6   &      0.0 \\ 
N4710    &     0.78   &  -0.27  &  3.02 &  2.11  &    1   &     +0.8 \\        
N4733    &     0.21   &   2.51  &  3.33 &  3.06  &    1.5 &     +0.7 \\         
N4753    &     0.21   &   1.98  &  4.23 &  3.51  &    3   &     +0.8 \\          
N4754    &     0.10   &   1.59  &  5.21 &  3.45  &    5   &     +0.7 \\         
N4762    &     0.26   &   0.75  &  4.90 &  3.02  &    3   &     +0.6 \\        
N5103    &     0.47   &   1.90  &  4.24 &  3.11  &    3   &     +0.7 \\     
N5342    &     0.16   &   1.19  &  4.96 &  2.70  &   15   &     +0.1 \\      
N5353    &     0.28   &   0.98  &  5.53 &  3.14  &   15   &     +0.3 \\         
N5355    &     0.33   &   5.23  &  2.26 &  2.14  &    1   &     +0.7 \\         
N5358    &     0.33   &   2.02  &  3.46 &  2.77  &    3   &     +0.2 \\        
N5422    &     0.60   &   1.29  &  4.80 &  3.16  &    7   &     +0.4 \\        
N5473    &     0      &   1.89  &  4.50 &  3.06  &    4   &     +0.4 \\                
N5485    &     0.23   &   1.61  &  4.84 &  2.93  &    6   &     +0.4 \\         
N5493    &     0.20   &   2.06  &  3.25 &  2.62  &    4   &      0.0 \\               
N5507    &     0.53   &   1.11  &  6.04 &  3.04  &    9   &     +0.5 \\      
N5574    &     0.37   &   3.94  &  2.48 &  2.52  &    1   &     +0.8 \\       
N5611    &     0.28   &   1.94  &  4.03 &  2.70  &    4   &     +0.2 \\      
N5631    &     0.44   &   1.79  &  4.74 &  3.23  &    3   &     +0.8 \\      
N5687    &     0.15   &   1.77  &  4.85 &  3.50  &    4   &     +0.8 \\     
N5770    &     0.27   &   2.19  &  4.15 &  3.14  &    2   &     +0.8 \\     
N5839    &     0.13   &   1.52  &  5.02 &  3.32  &    6   &     +0.5 \\     
N5854    &     0.21   &   2.41  &  2.98 &  2.58  &    2   &     +0.1 \\     
N5864    &     0.22   &   1.95  &  3.80 &  3.10  &    3   &     +0.5 \\      
N5869    &     0      &   1.70  &  5.19 &  3.18  &    5   &     +0.5 \\     
N6010    &     0.29   &   2.09  &  4.54 &  3.07  &    2   &     +0.7 \\      
N6017    &     1.13   &   1.83  &  3.56 &  2.60  &    2   &     +0.4 \\        
N6149    &     1.06   &   1.63  &  3.98 &  2.16  &    3   &     +0.1 \\       
N6278    &     0.56   &   1.06  &  4.91 &  3.10  &   14   &     +0.2 \\        
N6703    &     0.23   &   1.58  &  4.88 &  3.26  &    5   &     +0.5 \\        
N6798    &     1.25   &   1.59  &  4.65 &  3.49  &    2   &     +1.0 \\        
N7693    &     0.19   &   4.71  &  1.89 &  2.18  &    1   &     +0.8 \\        
N7710    &     0.69   &   1.52  &  3.43 &  2.89  &    5   &     +0.1 \\       
N7743    &     5.5    &   2.03  &  3.30 &  2.54  &    1   &     +1.0 \\       
 I560    &     3.17   &   1.97  &  2.48 &  2.22  &    1   &     +0.2  \\      
 I598    &     0.54   &   2.65  &  2.59 &  2.35  &    1.7 &     +0.1 \\        
 I719    &     0.39   &   1.68  &  3.45 &  2.77  &    6   &      0.0  \\      
 I782    &     0.35   &   2.92  &  3.11 &  2.44  &    1   &     +0.7 \\                  
U4551    &     0.23   &   1.65  &  4.16 &  2.86  &    8   &     +0.15 \\     
U6062    &     0.25   &   1.50  &  4.45 &  3.22  &    8   &     +0.3  \\     
U8876    &     0.11   &   1.73  &  3.98 &  2.64  &    9   &      0.0 \\     
U9519    &     0.38   &   3.04  &  2.77 &  2.36  &    1   &     +0.3 \\           
P28887   &     0.43   &   1.42  &  4.24 &  2.32  &   12   &     -0.15 \\     
P35754   &     0.78   &   1.97  &  3.75 &  2.32  &    2   &     +0.2 \\     
P42549   &     1.31   &   2.31  &  3.11 &  2.84  &    1   &     +0.8 \\      
P44433   &     0.34   &   1.92  &  4.13 &  2.76  &    3   &     +0.3 \\       
P50395   &     0.39   &   2.34  &  3.32 &  2.54  &    2   &     +0.15 \\        
P51753   &     0.88   &   1.65  &  3.52 &  2.59  &    4   &     +0.1 \\      
P54452   &     0.39   &   2.10  &  3.50 &  3.25  &    2   &     +0.5 \\        
\enddata
\tablenotetext{a}{Galaxy ID -- N=NGC, U=UGC, I=IC, P=PGC}
\end{deluxetable}

\begin{deluxetable}{lccccccccc}
\tablenum{2}
\tablewidth{16cm}
\tablewidth{0pt}
\tabletypesize{\small}
\tablecaption{The Lick indices and SSP-equivalent parameters for the bulges}
\tablehead{
\colhead{Galaxy\tablenotemark{a}}  &  
\colhead{$r_e$(bul), $^{\prime \prime}$} &
\colhead{Its source\tablenotemark{b}} &
\colhead{Eq. width of [OIII]$\lambda$5007} &
\colhead{H$\beta$}  &
\colhead{Mgb}  &
\colhead{Fe5270} &
\colhead{T, Gyr} &
\colhead{[Z/H], dex} &
\colhead{[Mg/Fe]}
}
\startdata
 N448 &  4.7  & 4 & 0.10 &  1.72 &  3.10  &  2.54  & 10 &  -0.2 &  +0.1 \\
 N502 &  3.5  & 8 & 0.20 &  1.62 &  3.79  &  2.91  &  8 &  +0.1 &  +0.1 \\
 N509 &  2.9  & 8 & 0.55 &  2.08 &  2.50  &  2.48  &  3 &  -0.1 &   0.0 \\
 N516 &  2.9  & 8 & 0.41 &  1.83 &  2.84  &  2.65  &  5 &  -0.1 &   0.0 \\
 N525 &  3.7  & 4 & 0.25 &  1.96 &  3.15  &  2.11  &  6 &  -0.2 &  +0.3 \\
N2577 &  4.0  & 4 & 0    &  1.55 &  3.99  &  2.63  & 14 &  -0.1 &  +0.3 \\
N2698 &  2.2  & 4 & 0.10 &  1.72 &  4.53  &  2.89  &  8 &  +0.2 &  +0.3 \\
N2852 &  5.2  & 4 & 0.61 &  1.51 &  3.39  &  2.02  &  9 &  -0.3 &  +0.4 \\
N2859 &  4.7  & 3 & 0.33 &  1.51 &  4.13  &  2.82  & 10 &  +0.1 &  +0.2 \\
N2880 &  6.4  & 1 & 0.16 &  1.86 &  3.35  &  2.57  &  7 &  -0.1 &  +0.15 \\
N2950 &  5.1  & 1 & 0.50 &  1.76 &  3.28  &  2.57  &  4 &   0.0 &  +0.1 \\
N2962 &  $3.5\pm 0.8$  & 3,7 & 0.63 &  1.22 &  4.21  &  2.72  & 12 &   0.0 &  +0.3 \\
N3230 &  4.2  & 5 & 0.13 &  1.20 &  4.11  &  2.64  & 15 &  -0.1 &  +0.3 \\
N3245 &  $3.5\pm 0.2$  & 1,5 & 0.15 &  1.42 &  3.88  &  2.65  & 15 &  -0.15 & +0.2 \\
N3248 &  4.4  & 3 & 1.03 &  1.84 &  2.76  &  2.37  &  3 &  -0.05 & +0.1 \\
N3301 &  $3.3\pm 0.4$  & 3,7 & 0.38 &  1.88 &  2.81  &  2.56  &  5 &  -0.2  &  0.0 \\
N3412 &  8.1  & 7 & 0.22 &  1.85 &  3.03  &  2.50  &  6 &  -0.15 & +0.1 \\
N3599 &  3.4  & 3 & 1.35 &  1.88 &  2.72  &  2.64  &  2 &  +0.1 &   0.0 \\
N3607 &  $7.8\pm 0.3$  & 1,5 & 0.43 &  1.27 &  4.04  &  2.64  & 15 &  -0.1 &  +0.3 \\
N3610 &  3.9  & 4 & 0.37 &  1.97 &  3.40  &  2.57  &  4 &  +0.05 & +0.15 \\
N3613 &  7.6  & 4 & 0.12 &  1.67 &  4.08  &  2.64  &  9 &   0.0 &  +0.3 \\
N3626 &  $2.5\pm 0.1$  & 3,5 & 0.40 &  2.53 &  2.43  &  2.37  &  2 &   0.0 &   0.0 \\
N3630 &  $3.7\pm 0.6$  & 4,7 &  0.14 &  1.66 &  3.74  &  2.68  & 10 &  -0.1 &  +0.15 \\
N3665 & 14.4  & 1 & 0.16 &  1.68 &  3.49  &  2.35  & 10 &  -0.2 &  +0.3 \\
N3674 &  2.9  & 4 & 0.41 &  1.40 &  3.88  &  2.53  & 13 &  -0.2 &  +0.3 \\
N3757 &  2.1  & 3 & 0.17 &  1.78 &  3.62  &  2.70  &  7 &   0.0 &  +0.15 \\
N3838 &  2.1  & 4 & 0.34 &  1.98 &  3.41  &  2.52  &  4 &   0.0 &  +0.2 \\
N3941 &  $2.9\pm 0.5$  & 2,3 &  0.69 &  1.50 &  3.77  &  2.79  &  5 &  +0.1 &  +0.15 \\
N3945 &  8.5  & 1 & 0.33 &  1.32 &  3.99  &  2.80  & 14 &  -0.1 &  +0.15 \\
N4036 &  8.9  &  5 & 0.47 &  1.37 &  3.86  &  2.64  & 12 &  -0.1 &  +0.2 \\
N4111 &  3.0  & 3 & 0.40 &  1.68 &  3.55  &  2.74  &  5  & +0.1 &  +0.1 \\
N4124 &  8.0  & 4 & 0.42 &  2.60 &  2.62  &  2.39  & 1.5 & +0.1 &   0.0 \\
N4179 &  8.1  & 4 & 0.18 &  1.81 &  3.48  &  2.52  &  7  & -0.1 &  +0.2 \\
N4233 &  $4.6\pm 0.2$  & 4,5 & 0.54 &  1.01 &  4.06  &  2.58  & 15  & -0.1 &  +0.3 \\
N4267 &  4.4  &  1 & 0.12 &  1.55 &  4.21  &  2.65  & 12  &  0.0 &  +0.3 \\
N4324 &  7.8  & 3 &  1.07 &  1.49 &  3.06  &  2.59  &  4  &  0.0 &   0.0 \\
N4340 &  4.4  & 2 & 0.23 &  1.72 &  3.33  &  2.99  &  7  &  0.0 &   0.0 \\
N4342 &  2.6  & 6 & 0.1  &  1.37 &  4.69  &  2.90  & 15  & +0.1 &  +0.3 \\
N4346 &  $3.8\pm 0.4$  & 3,4 & 0.34 &  1.90 &  3.38  &  2.70  &  4  & +0.1 &  +0.1 \\
N4350 &  3.0  & 4 & 0.12 &  1.32 &  4.62  &  2.87  & 15  & +0.1 &  +0.3 \\
N4371 &  7.5  & 3 & 0.13 &  1.61 &  3.83  &  2.63  & 11  & -0.1 &  +0.2 \\
N4377 &  $3.4\pm 0.9$  & 5,7 &  0.12 &  1.65 &  3.35  &  2.51  & 11  & -0.2 &  +0.15 \\
N4417 &  5.3  & 4 & 0.17 &  1.64 &  3.48  &  2.44  & 11  & -0.2 &  +0.2 \\
N4429 & 10.7  & 5 & 0.13 &  1.74 &  4.40  &  2.62  &  7  & +0.1 &  +0.3 \\
N4434 &  4.7  & 4 & 0.24 &  1.59 &  3.36  &  2.59  & 10  & -0.15 & +0.1 \\
N4442 &  9.2  & 3 & 0.10 &  1.36 &  4.23  &  2.65  & 15  & -0.1 &  +0.3 \\
N4461 &  4.4  & 3 & 0.19 &  1.69 &  3.91  &  2.77  &  7  & +0.1 &  +0.15 \\
N4474 &  3.5  & 4 & 0.16 &  1.92 &  3.32  &  2.52  &  5  & -0.1 &  +0.15 \\
N4476 &  8.2  & 6 & 0.52 &  2.02 &  2.12  &  1.72  &  5  & -0.4 &  +0.15 \\
N4483 &  3.7  & 7 & 0.17 &  1.93 &  3.53  &  2.50  &  6  & -0.05 & +0.2 \\
N4503 &  9.7  & 3 & 0.21 &  1.60 &  3.94  &  2.67  & 10  &  0.0 &  +0.2 \\
N4528 &  3.7  & 5 & 0.16 &  1.97 &  2.96  &  2.19  &  7  & -0.25 & +0.2 \\
N4578 & $10.3\pm 3.9$  & 4,5 & 0.42 &  1.94 &  3.36  &  2.49  &  4  &  0.0 &  +0.15 \\
N4596 &  2.8  & 2 &  0.31 &  1.17 &  4.23  &  2.80  & 15  & -0.05 & +0.2 \\
N4608 &  3.3  & 2 & 0.15 &  1.64 &  3.85  &  2.69  & 10  &  0.0 &  +0.2 \\
N4623 & 11.9  & 4 & 0.26 &  2.02 &  3.00  &  2.49  &  4  & -0.1 &  +0.1 \\
N4643 &  $4.2\pm 1.0$  & 2,3 & 0.26 &  1.45 &  3.86  &  2.76  & 12  & -0.1 &  +0.15 \\
N4733 &  5.7  & 3 & 0.25 &  2.06 &  2.82  &  2.51  &  5  & -0.15 &  0.0 \\
N4754 &  $4.4\pm 1.6$  & 1,3 & 0.   &  1.39 &  3.93  &  2.79  & 15  & -0.1 &  +0.15 \\
N4762 &  3.1  & 4 & 0.24 &  1.77 &  4.13  &  2.82  &  5  & +0.2 &  +0.2 \\
N5103 &  1.9  & 4 & 0.44 &  1.64 &  3.51  &  2.65  &  7  & -0.1 &  +0.15 \\
N5342 &  1.7  & 4 & 0.16 &  1.27 &  4.01  &  2.25  & 15  & -0.2 &  +0.4 \\
N5353 &  5.4  & 7 & 0    &  1.32 &  4.82  &  3.15  & 15  & +0.25 & +0.2 \\
N5355 &  7.4  &  3 & 0.40 &  2.11 &  2.43  &  2.22  &  4  & -0.2  &  0.0 \\ 
N5358 &  2.1  & 4 & 0.23 &  1.96 &  3.13  &  2.59  &  4  &  0.0 &  +0.1 \\
N5422 &  $6.2\pm 2.0$  & 3,4 & 0.53 &  1.70 &  3.78  &  2.32  &  6  & -0.1 &  +0.4 \\
N5473 &  $2.7\pm 0.1$  & 1,3 & 0.   &  1.73 &  4.26  &  2.74  &  9  & +0.1 &  +0.4 \\
N5485 & $10.6\pm 1.6$  & 1,5 & 0.16 &  1.67 &  4.10  &  2.54  &  9  &  0.0 &  +0.3 \\
N5493 &  5.9  &  1 & 0.14 &  1.73 &  2.36  &  2.17  & 13  & -0.4 &   0.0 \\
N5507 &  $3.0\pm 0.1$  & 3,4 &  0.49 &  1.21 &  4.32  &  2.56  & 15  & -0.1 &  +0.3 \\
N5574 & 10.3  & 3 & 0.31 &  2.59 &  2.69  &  1.94  &  2  & -0.2 &  +0.3 \\
N5611 &  2.1  & -- & 0.18 &  1.62 &  3.17  &  2.32  & 12  & -0.3 &  +0.2 \\
N5631 &  6.5  & 1 & 0.34 &  1.60 &  3.28  &  2.47  &  9  & -0.15 & +0.15 \\
N5687 &  $5.7\pm 1.8$ & 4,5 &  0    &  1.55 &  3.97  &  2.75  & 14  & -0.1 &  +0.2 \\
N5854 &  4.9  & 3 & 0.45 &  2.46 &  2.94  &  2.56  & 1.5 & +0.2 &   0.0 \\
N5864 &  2.5  & 3 & 0.21 &  1.85 &  3.38  &  2.65  &  5  &  0.0 &  +0.1 \\
N6017 &  7.6  & 4 & 0.455 & 1.94 &  2.17  &  1.88  &  5  & -0.4 &   0.0 \\
N6149 &  3.2  & 4 & 0.60 &  1.31 &  2.98  &  2.30  & 15  & -0.4 &  +0.2 \\
N6278 &  $2.9\pm 0.6$ & 3,4,5 &  0.25 &  1.22 &  4.60  &  3.12  & 15  & +0.2 &  +0.2 \\
N6703 &  4.2  &  1 & 0.22 &  1.44 &  3.94  &  2.99  & 11  & +0.1 &  +0.1 \\
N7710 &  3.0  & -- & 0.30 &  1.26 &  2.94  &  2.27  & $>8$  & $<-0.4$ &  +0.2 \\
N7743 &  $2.7\pm 1.1$  & 1,3 & 1.05 &  1.99 &  2.59  &  2.77  &  2  & +0.1 &   0.0 \\
 I560 &  2.9  & 4 & 1.62 &  2.01 &  2.56  &  2.02  &  2  & -0.2 &  +0.2 \\
U4551 &  3.4  & 3 & 0.42 &  1.48 &  3.70  &  2.58  & 10  & -0.1 &  +0.2 \\
P28887 & 3.7  & 4 & 0.37 &  1.21 &  2.54  &  1.74  & $>8$  & $<-0.4$ &  +0.3 \\
P35754 & 3.1  & 4 & 0.78 &  1.44 &  2.94  &  1.99  &  9  & -0.35 & +0.3 \\
\enddata
\tablenotetext{a}{Galaxy ID -- N=NGC, U=UGC, P=PGC, I=IC}
\tablenotetext{b}{The sources of the $r_e(bul)$ are numerated as follows:\\
1 -- \citet{lauri10}\\
2 -- \citet{lauri05}\\
3 -- \citet{s4gdecomp}\\
4 -- \citet{atlas3d_17}\\
5 -- \citet{mendez}\\
6 -- \citet{donofrio}\\
7 -- \citet{baggett}\\
8 -- \citet{n524gr}\\
}
\end{deluxetable}

\begin{figure}
\plottwo{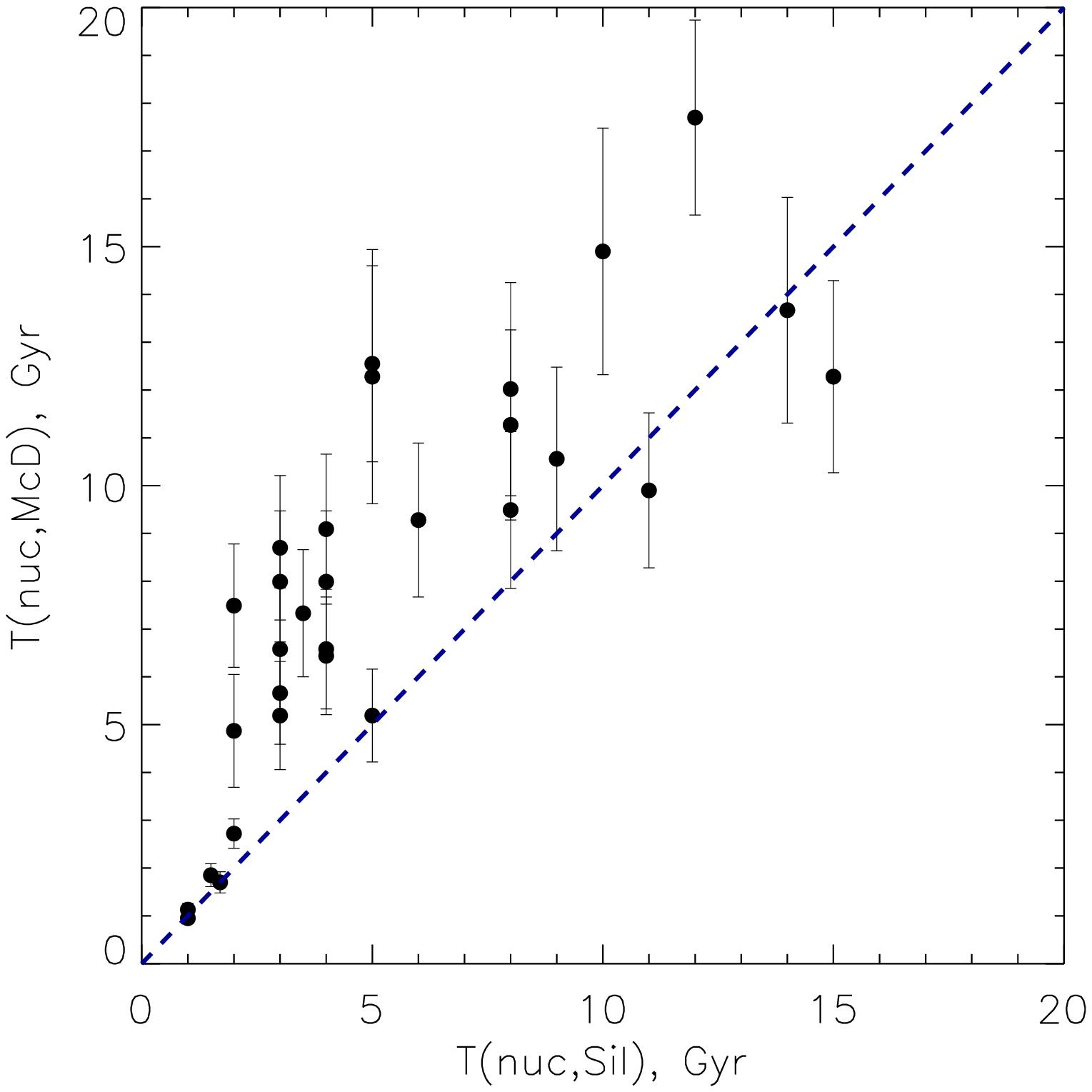}{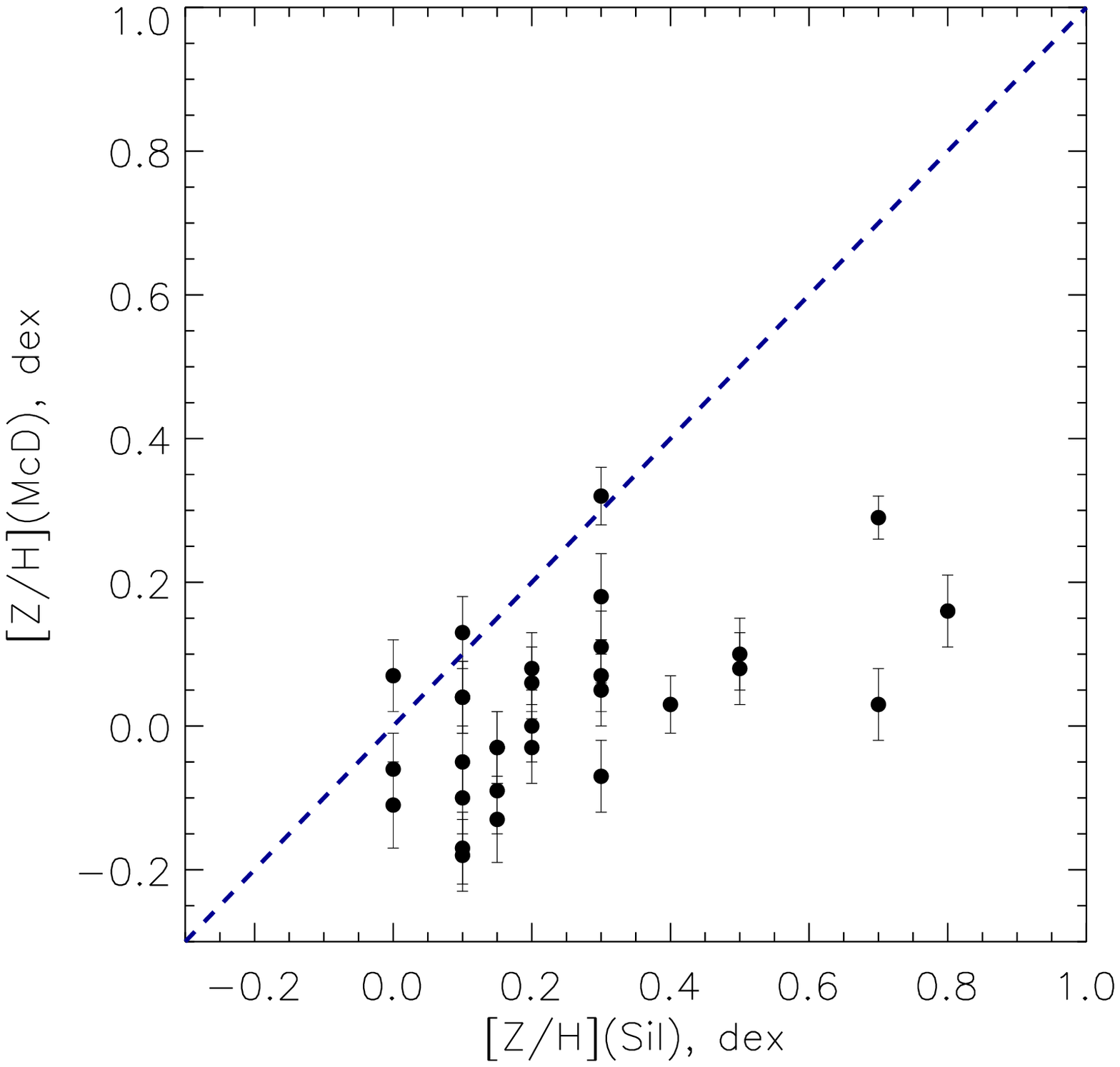}
\caption{The comparison of the nuclear stellar population parameters -- the SSP-equivalent ages and
metallicities -- derived through the Lick indices in this work, with those obtained by \citet{atlas3d_30}, 
for the subsample with $R_e/8$ less than 1.5 arcsec.}
\label{vsmcdermid}
\end{figure}

The team of the ATLAS-3D survey made its own investigation of the stellar population properties
of the same targets \citep{atlas3d_30}. However they considered the stellar population properties
within apertures centered onto the nuclei, with radii taken at fixed fractions of the {\it totally measured} 
(integrated) effective radius -- $R_e/8$, $R_e/2$, $R_e$.
Meantime the lenticular galaxies that constitute the dominant part of the ATLAS-3D survey
consist of at least {\it two} large-scale components with different properties and evolution,
bulges and disks, having very different ratios of {\it their own} effective radii. So sometimes
the characteristics measured by the team of ATLAS-3D within the $R_e/2$ and $R_e$ apertures
relate to the bulges, sometimes -- to the disks, without specifying their attribution by \citet{atlas3d_30}. 
My investigations of the radially resolved properties of stellar populations in S0 galaxies have revealed
quite different evolution of the S0 galaxy constituents -- stellar nuclei, bulges, and the disks
\citep{s0mpfs,silsymp245,s0disks}, -- so I prefer to analyze {\it separately} the stellar nuclei and the bulges 
when I operate with the SAURON data. I can compare directly my estimates with the results of \citet{atlas3d_30}
only in one case: if I take my measurements for the {\it nuclei} and the measurements by \citet{atlas3d_30} -- for 
$R_e/8$, after selecting the subsample for comparison by putting a limit onto the integrated effective radius
presented in \citet{atlas3d_1}: $R_e<12^{\prime \prime}$. This limit allows to consider the $R_e/8$ apertures for the
comparison galaxies as close to the size of the resolved spatial element (seeing), 
given $R_e/8 < 1.5^{\prime \prime}$. The comparison of the nuclear stellar population properties is presented 
in Fig.~\ref{vsmcdermid}.
The ages estimates are well correlated though for intermediate ages, $T=3-7$~Gyr, a shift by about 4~Gyr 
can be noted, the ages by \citet{atlas3d_30} looking older. Subsequently, the metallicity estimates are
also correlated, but my estimates are higher by some 0.2 dex; three galaxies with [Z/H]$>0.6$ are
strongly off, but such high metallicities are absent in the model set by \citet{thomod} and are perhaps
improperly extrapolated. Taking into account the fact that the indices were confronted by me and by \citet{atlas3d_30} 
to quite different SSP-models, and that the correction of the H$\beta$ indices for the emission was made 
through quite different approaches, the divergence of the central stellar population parameters obtained so far may
be attributed to this fact. The difference in both models and H$\beta$ correction for the emission contamination
is significant. Concerning the former issue, I use the models by \citet{thomod} which are based on the
classical fitting functions for stellar Lick indices from \citet{woretal}, with the inserted
dependence on abundance ratios from \citet{tribell}, and on the fuel consumption theorem approach \citep{renbuz}; 
while \citet{atlas3d_30} use the \citet{schiavon} models, which are based on the fitting functions renewed by \citet{schiavon}  
and on the theoretical isochrone (so called `conventional') approach. Concerning the index H$\beta$ correction, 
I use the empirical calibration of the H$\beta$ emission line through the [O\iii]$\lambda$5007 emission line from 
\citet{trager00}, while \citet{atlas3d_30} used the approach proposed by the SAURON team \citep{sarzi06} which 
allows to derive full emission-line spectra by subtracting the preliminary fitted full spectrum of the stellar populations.

\begin{figure}
\plottwo{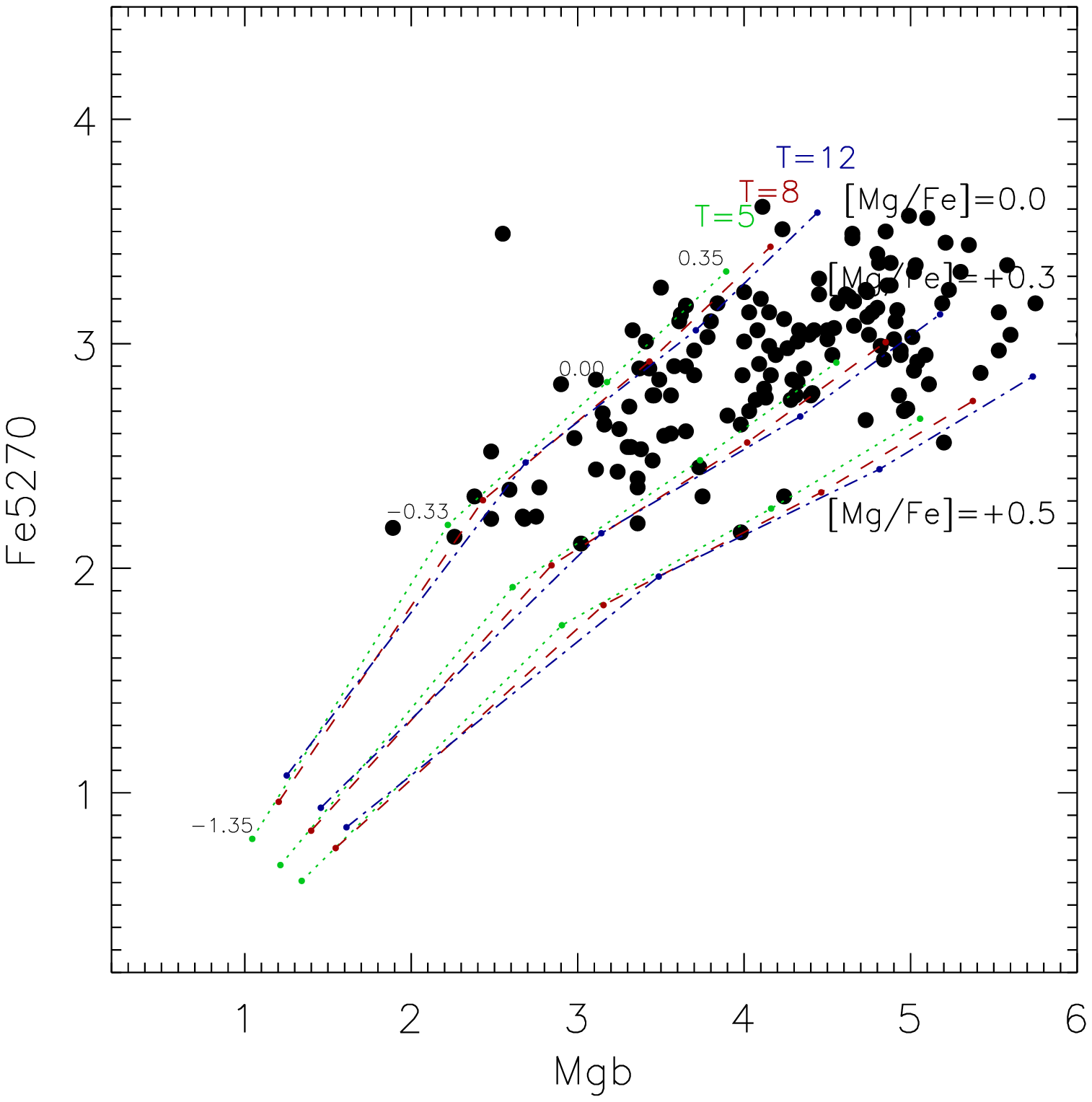}{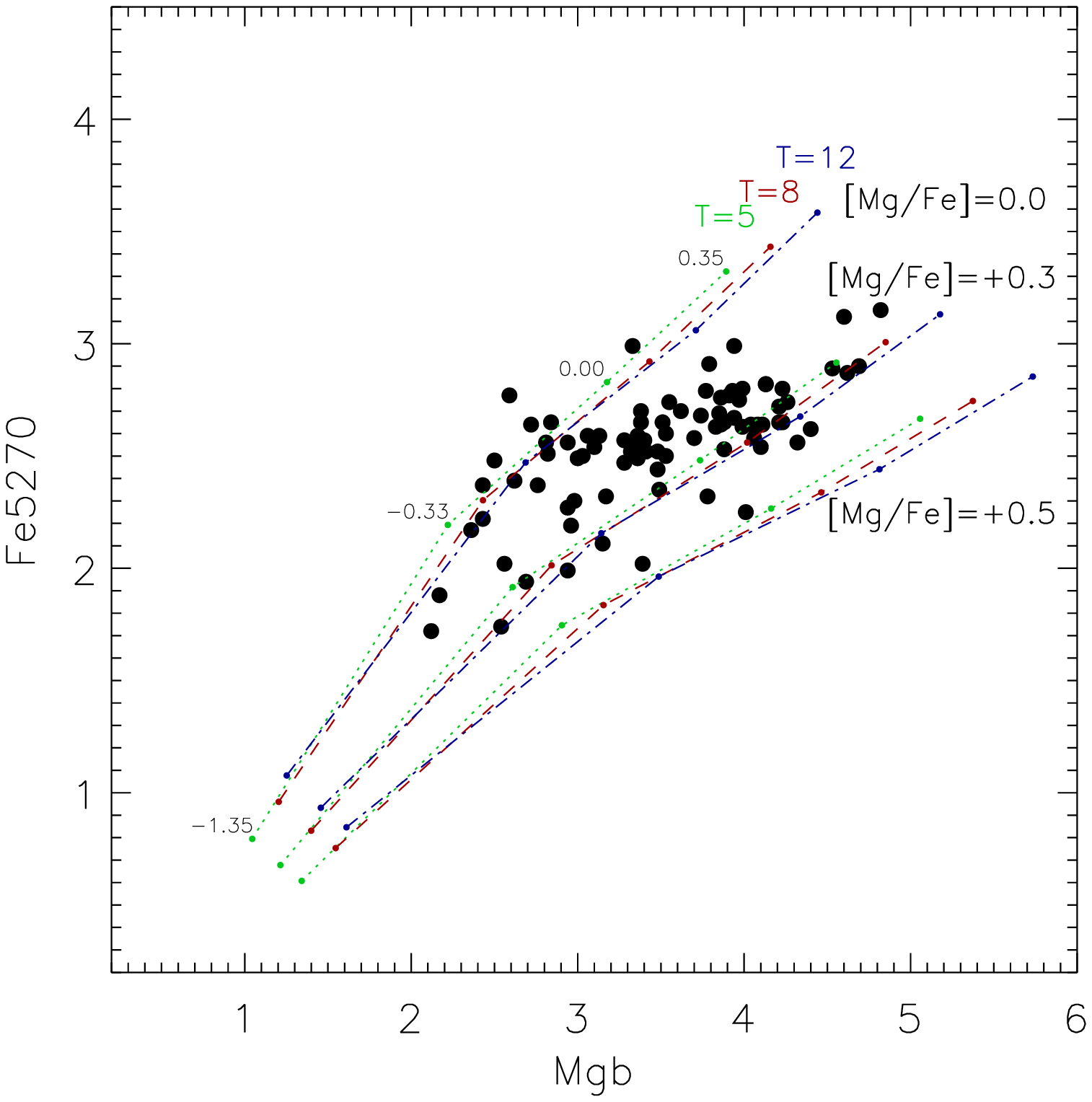}
\caption{The Mgb versus Fe5270 diagrams for the nuclei ({\it the left plot}) and for the bulges
at one {\bf bulge} effective radius ({\it the right plot}). The points are the galaxies of the sample, the lines
refer to the models: the simple stellar population models by \citet{thomod} for three different
magnesium-to-iron ratios (0.0, $+0.3$, and $+0.5$) and three different ages (5, 8, and 12~Gyr) are plotted 
as a reference frame. The small signs along the model curves mark the metallicities of $+0.35$, 0.00,
--0.33, and --1.35, if one takes the signs from right to left.}
\label{mgfe}
\end{figure}

\begin{figure}
\begin{tabular}{c c}
 \includegraphics[width=4cm]{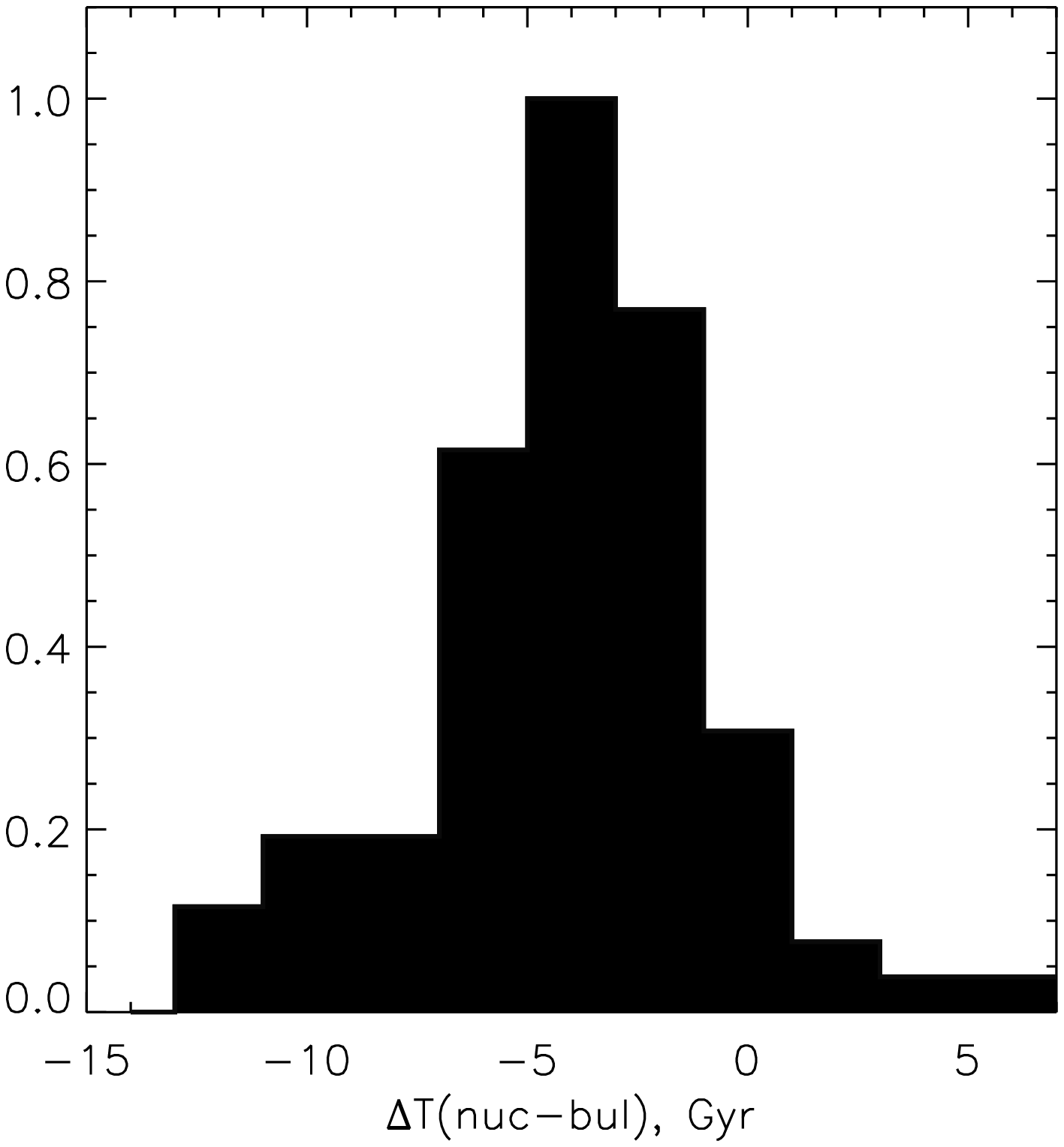} &
 \includegraphics[width=4cm]{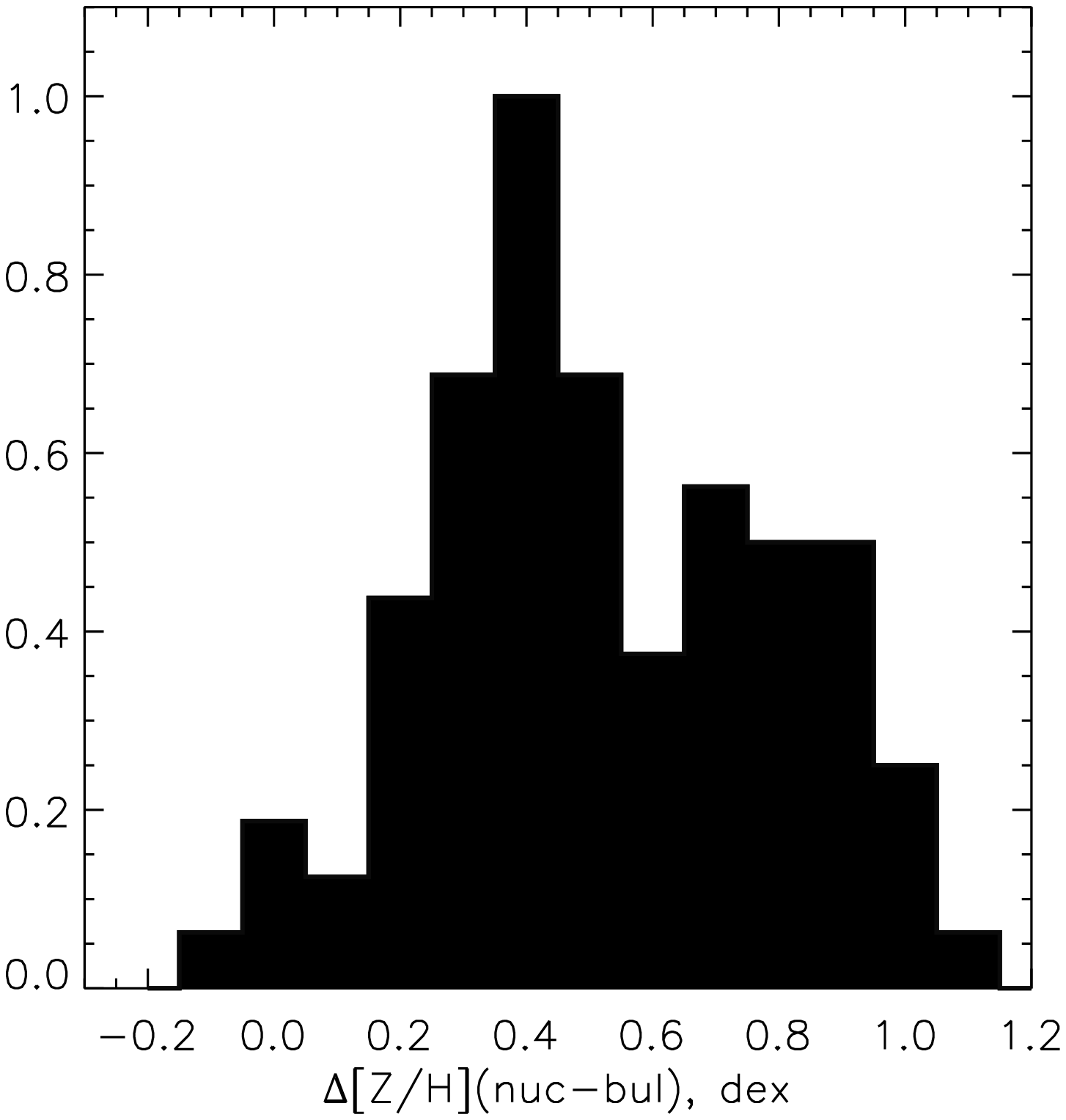} \\
\end{tabular}
\caption{The distributions of the stellar population parameter differences `nucleus minus bulge':
{\it left} -- the age difference between the nuclei and bulges,  {\it right} -- the 
metallicity difference between the nuclei and bulges. The histograms are normalized to unity
at the maximum.}
\label{delta_age_z_histo}
\end{figure}

\begin{figure}
\begin{tabular}{c c}
 \includegraphics[width=4cm]{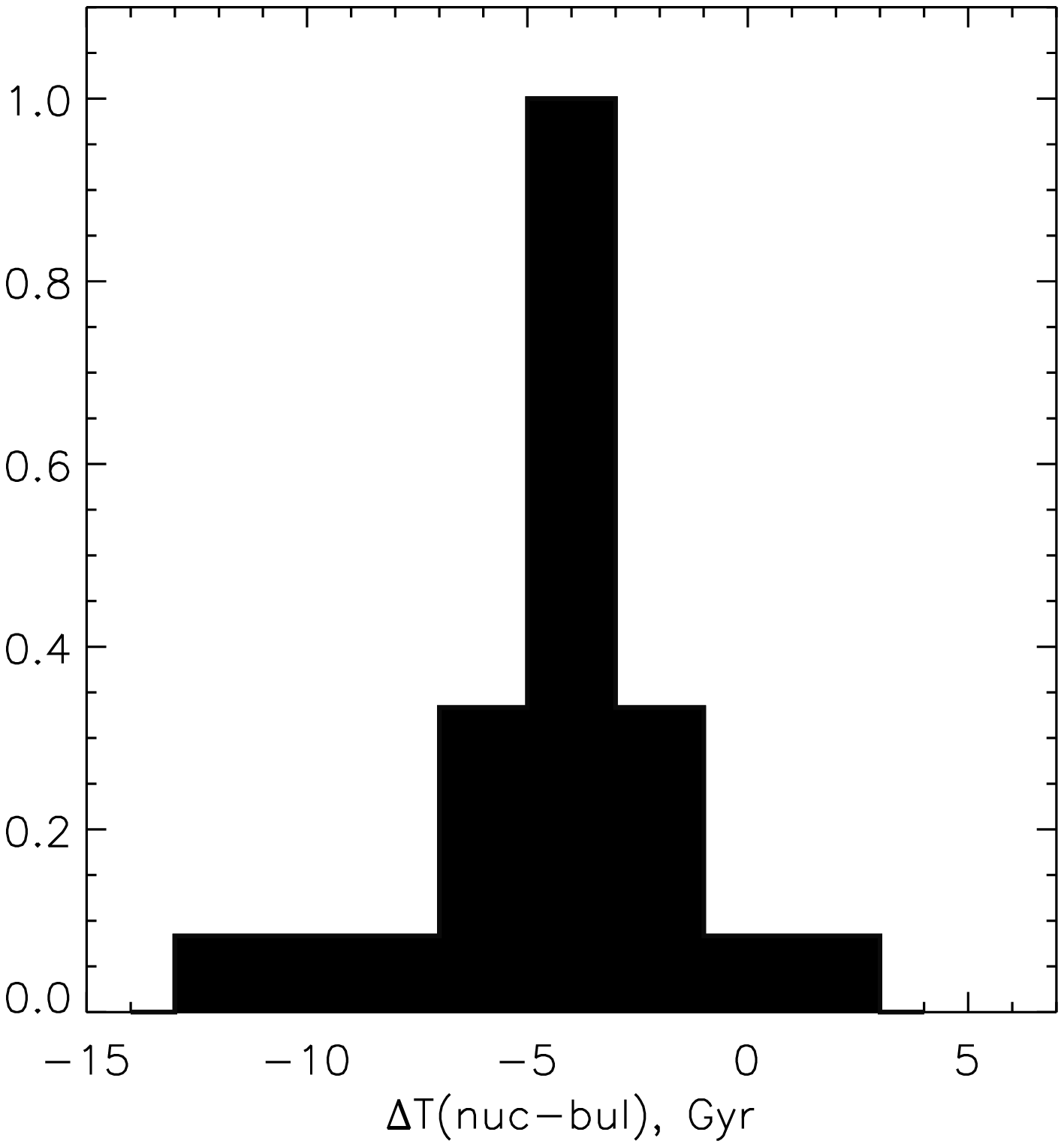} &
 \includegraphics[width=4cm]{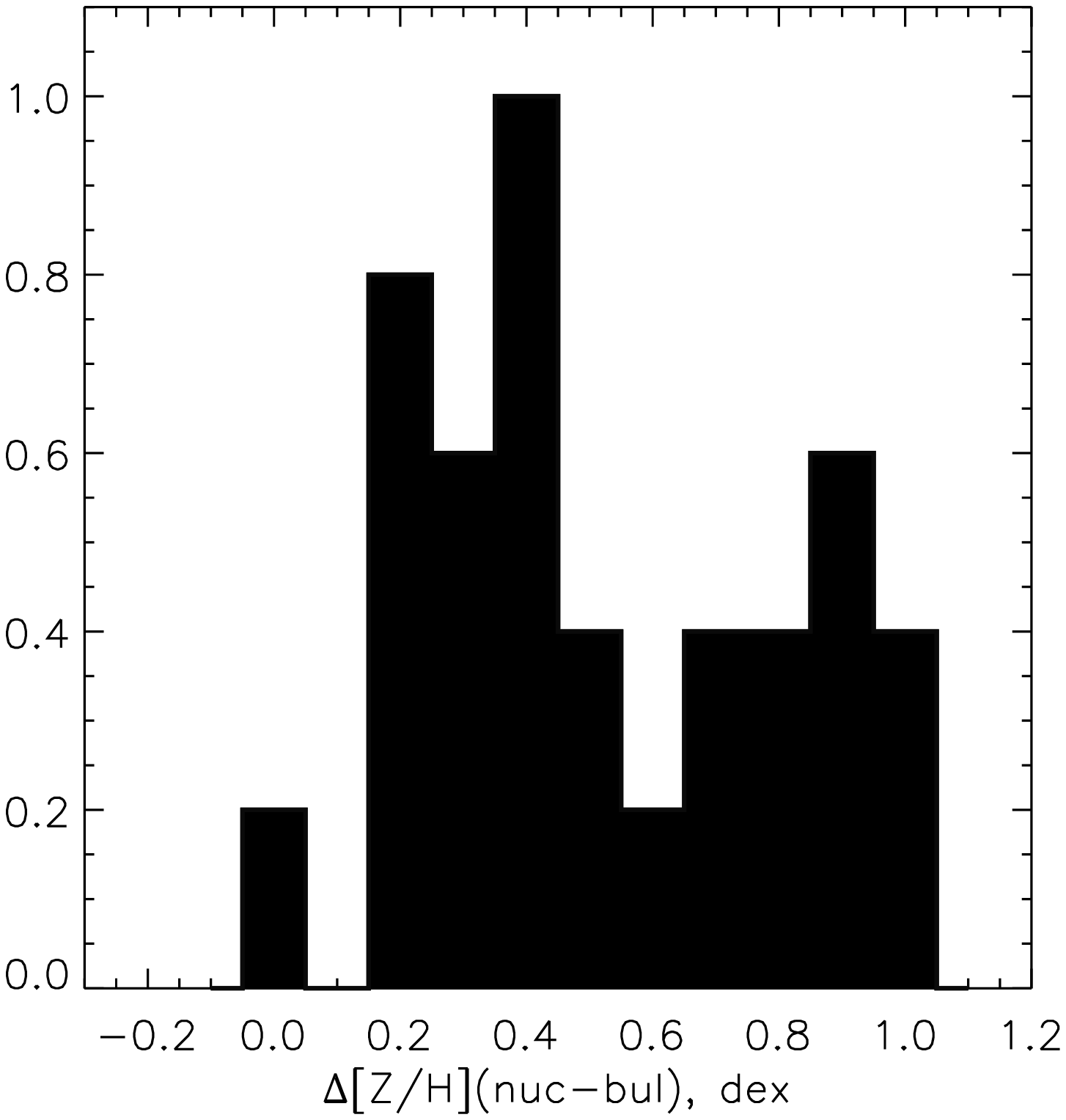} \\
 \includegraphics[width=4cm]{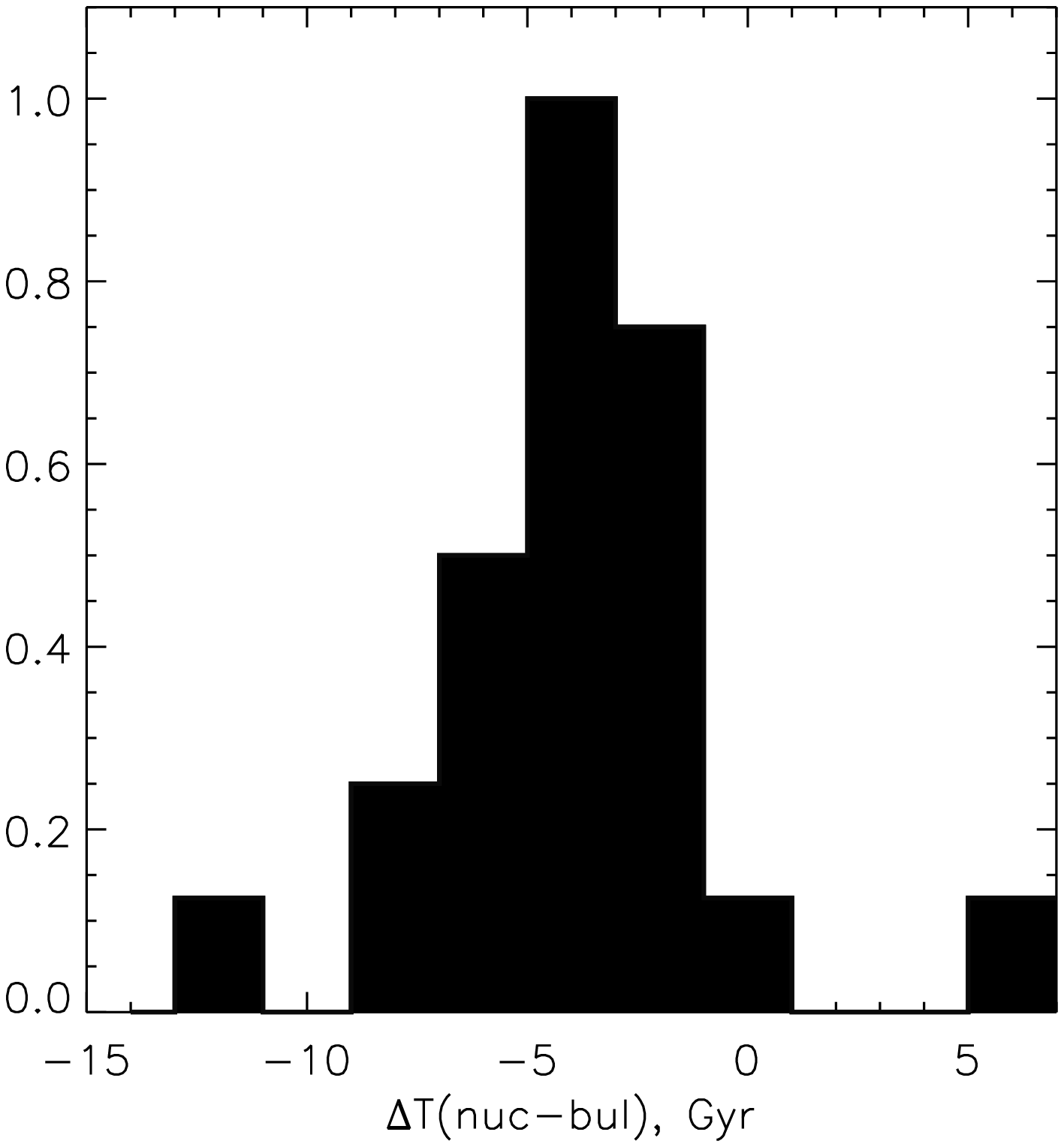} &
 \includegraphics[width=4cm]{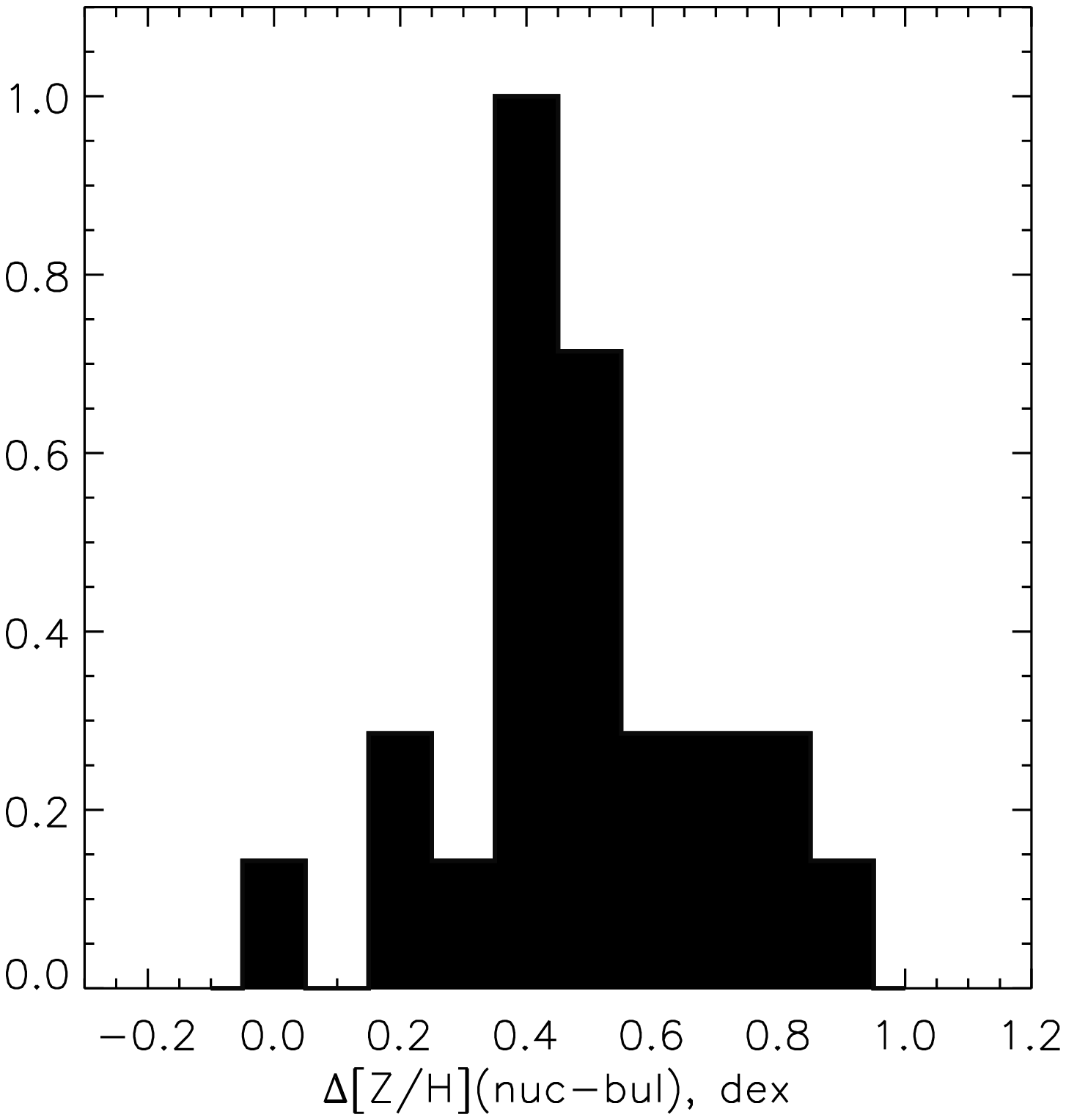} \\
\end{tabular}
\caption{The same as in Fig.~6 but with separation according to the environment density:
the upper row presents the results for the Virgo S0s, the bottom row -- for the galaxies
within environment sparser than the median one for the whole sample. The data on the environment
density are taken from \citet{atlas3d_7}. The histograms are normalized to unity
at the maximum.}
\label{delta_age_z_histo_env}
\end{figure}

As for the parameters of the stellar populations in the bulges (Table~2), they are obtained not for the whole sample but
only for the galaxies whose $r_e(bul)$ are within the field of view of the SAURON, and also can be
resolved with the typical seeing quality of $1.5^{\prime \prime}$; in other words, to study the bulge stellar
population properties, I have selected from the ATLAS-3D sample only the
S0 galaxies with $1.5^{\prime \prime} < r_e(bul) < 15^{\prime \prime}$.
The team of ATLAS-3D had published their own decomposition of the galactic surface brightness
profiles into the sums of a Sersic bulge and an exponential disk \citep{atlas3d_17}. But the decomposition was made only
for a part of the sample, and by analyzing only one-dimensional profiles derived from the SDSS data. 
They failed to separate two large-scale components in some obvious cases of S0 galaxies
with noticeable disks and bulges (NGC~3607, NGC~3665, NGC~4036, NGC~5355). To estimate the bulge effective radii
in the small S0 galaxies NGC~5611 and NGC~7710 where the analysis of \citet{atlas3d_17} has not revealed any
disks, I have undertaken my own decomposition of the SDSS one-dimensional surface-brightness profiles and have
succeeded to derive the parameters separately for the bulges and for the disks. To involve more certain data 
on the bulge's effective radii, I have searched for results of 2D decomposition, by considering the decompositions made in the 
near-infrared light as the better ones. Also, other literature sources were searched for the galaxies which were not
involved into photometric surveys dedicated to the 2D surface-brightness decomposition. Finally I have collected 
the bulge's effective radius measurements for the S0 galaxies of the ATLAS-3D sample from 8 different sources listed
in the Table~2. When a galaxy was investigated in several works, and the derived $r_e(bul)$ were
close, I took the averaged value which is given in the Table~2 with a corresponding estimate of the $r_e(bul)$ scatter. 
When all the results for $r_e(bul)$ were strongly different, I took the estimates obtained by the 2D NIR decomposition --
from \citet{lauri10} or from \citet{s4gdecomp}. The adopted values of  $r_e(bul)$
are listed in the Table~2 with the corresponding references to their sources.

Figures~5--7 demonstrate some statistics on the stellar population properties obtained. 
The magnesium-to-iron ratios (Fig.~\ref{mgfe}) are mostly between [Mg/Fe]$=0$ and 
[Mg/Fe]$=+0.3$ both for the nuclei and for the bulges; however the whole distribution of the bulges at the
(Mgb, Fe5270) diagram is shifted to lower metallicities with respect to that of the nuclei. As I have noted
earlier more than once -- see e.g. \citet{decnuc,s0mpfs,silsymp245} -- the stellar nuclei are chemically and evolutionary
decoupled with respect to the bulges, they are not simply the central points (R=0) of the bulges. 
In Fig.~\ref{delta_age_z_histo} one can see that the nuclei are typically younger and more metal-rich than the bulges. 
Earlier \citep{silsymp245} I have estimated the fraction of chemically distinct nuclei in nearby S0s as a half 
of the sample; now, with the volume-limited sample of the ATLAS-3D, I increase this estimate to at least 84\%.
In Fig.~\ref{delta_age_z_histo} the age difference distribution has a maximum at --4~Gyr, and the metallicity
difference distribution has two peaks, at $+0.4$ dex and at $+0.8$ dex. I have divided the whole sample into two
subsamples according to the local environment density. The local environment densities for the galaxies of the
ATLAS-3D sample have been estimated by \citet{atlas3d_7}, and it is from this paper that I have taken 
the environment density estimators $\Sigma _{10}$ for every galaxy.  \citet{atlas3d_7} noted that a condition 
$\log \Sigma _{10} > 0.4$ separates well the Virgo members from other galaxies. So I have taken two subsamples, with
$\log \Sigma _{10} > 0.4$ (`cluster and rich-group members') and with $\log \Sigma _{10} < -0.65$ (`isolated galaxies'), 
to plot the distributions of the age and metallicity differences betwen the nuclei and the bulges 
in Fig.~\ref{delta_age_z_histo_env}. It becomes clear (Fig.~\ref{delta_age_z_histo_env}) 
that the second peak in the metallicity difference distribution is due completely to the Virgo members,
while the broadening of the age difference distribution is produced by the galaxies in low-density
environments. Evident dependence of the evolution paths of S0 galaxies on the environment density is implied
by these results.

\section{Newly discovered inner polar disks, and statistics of inner polar disks over the volume-limited sample of nearby S0s}

In the ATLAS-3D volume-limited sample of nearby S0s, I have found seven new cases of polar 
gas rotation; in 5 of them the gas velocity fields have appeared to be extended enough -- at least 5--10 spaxels
from the center, -- and the main stellar 
disks are not strictly edge-on, to give a possibility to apply to them the tilted-ring DETKA software. 
The results, namely, the orientation angles of the rotation planes 
for the stellar and for the gaseous components as well as the estimated mutual inclinations of the gaseous
and stellar rotation planes are presented in the Table~3. In three galaxies -- NGC~2962,
NGC~3648, and NGC~4690 -- I see warps of the inner gaseous disks or even several nested gaseous rings with
slightly different orientations; for these galaxies I give the results in two separate radius ranges.

\setcounter{table}{2}

\begin{table*}
\caption[] {Orientations of the rotation planes for the stars and ionized gas in 5 S0 galaxies}
\begin{flushleft}
\begin{tabular}{lrrcrrc}
\hline\noalign{\smallskip}
NGC & PA$_{kin}$(stars), deg & $i_{kin}$(stars),deg & Radius (gas), arcsec 
&  PA$_{kin}$(gas), deg &  $i_{kin}$(gas), deg & $\Delta \psi$, deg \\
\hline
2962 & $7\pm 1.5$ & $47\pm 3$ & 6--11 & $80\pm 3$ & $63.5\pm 2.5$ & 60 or 83.5 \\
2962 & $7\pm 1.5$ & $47\pm 3$ & 13--18 & $51.5\pm 1.5$ & $78\pm 1$ & 50 or 111 \\
3499 & $38\pm 4$ & $48\pm 8$ &  5--11 & $133\pm 3$ & $62\pm 5.5$ & 75 or 68 \\
3648 &  $253\pm 3$ & $54\pm 7$ & 1--4  & $282\pm 4$ & $66\pm 3$ & 28 or 114 \\
3648 &  $253\pm 3$ & $54\pm 7$ & 5--8  & $332\pm 3$ & $67\pm 1$ & 68 or 85 \\
4690 & $332\pm 7$ & $18\pm 10$ & 3--6 & $169\pm 2$ & $71\pm 4$ & 88 or 54 \\
4690 & $332\pm 7$ & $18\pm 10$ & 7--8 & $225\pm 8$  & $65\pm 13$ & 71 or 61 \\
5507 & $63\pm 3$ & $64\pm 5$ & 2--6 & $138\pm 2$ & $64\pm 4$ & 66 or 91 \\
\hline
\end{tabular}
\end{flushleft}
\end{table*}

\begin{figure*}
\resizebox{\hsize}{!}{\includegraphics{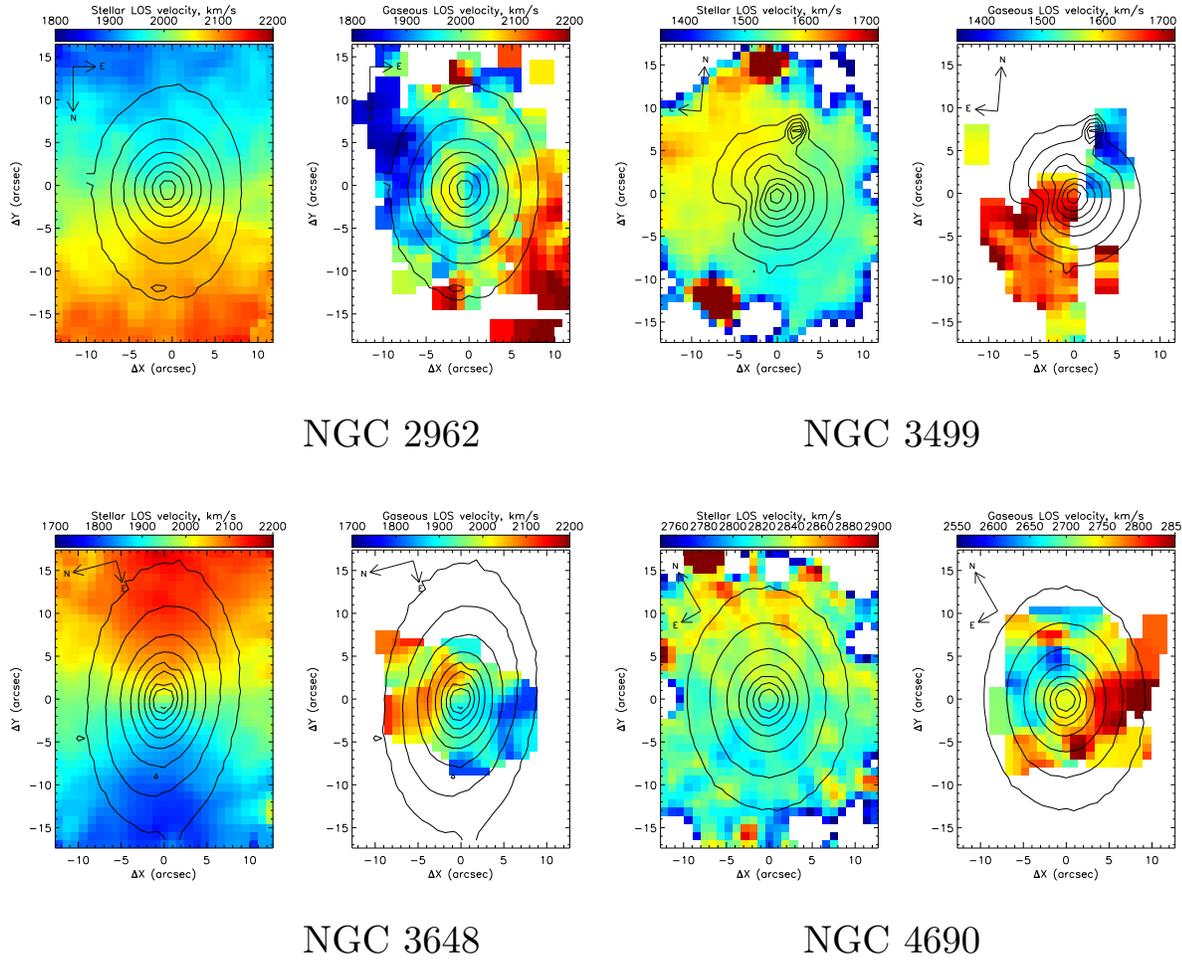}}
\caption{Line-of-sight velocity fields for the stellar (left plot in each pair) and ionized-gas
(right plot in each pair) components in 4 lenticular galaxies which have been constructed by using 
the SAURON panoramic spectral data. To derive the ionized-gas velocity fields, the measurements of the
[O\iii]$\lambda$5007 emission line were used. The green continuum isophotes, $\lambda=5100$~\AA, are overplotted.}
\label{velfields}
\end{figure*}

\begin{figure*}
\resizebox{\hsize}{!}{\includegraphics{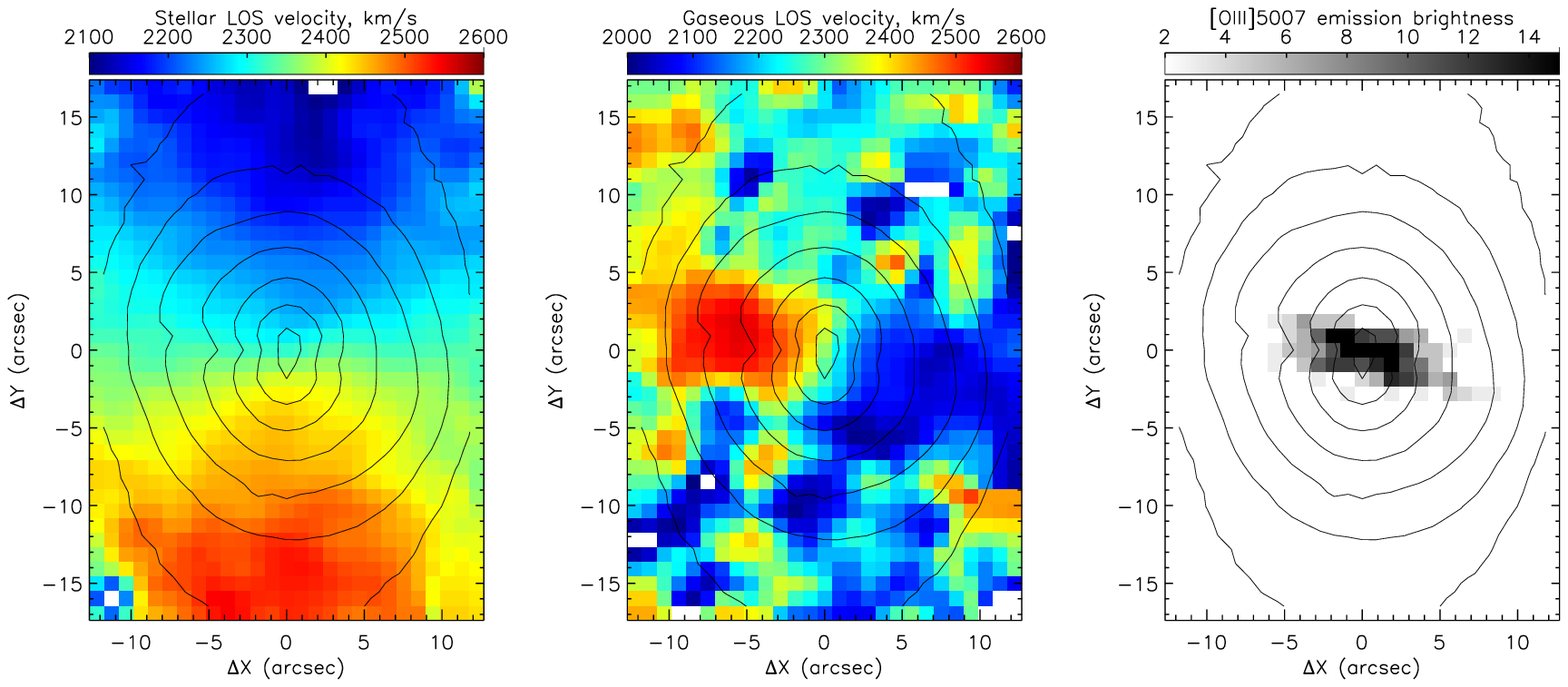}}
\resizebox{\hsize}{!}{\includegraphics{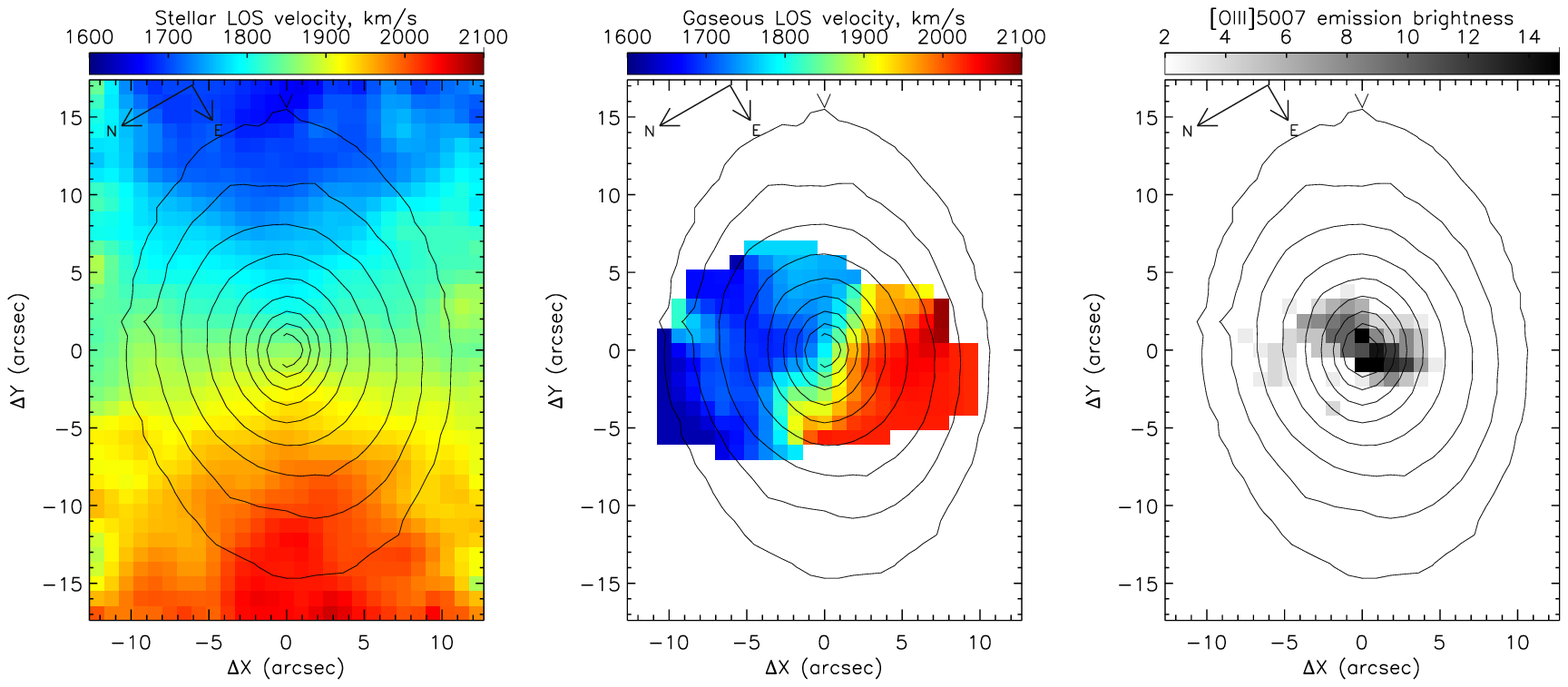}}
\caption{The stellar and gaseous velocity fields and the [O\iii] emission-line intensity distributions 
in NGC~4233 (up) and NGC 5507 (bottom). The green continuum isophotes, $\lambda=5100$~\AA, are overplotted.}
\label{emismaps}
\end{figure*}

\begin{figure*}
\begin{tabular}{c c}
 \includegraphics[width=8cm]{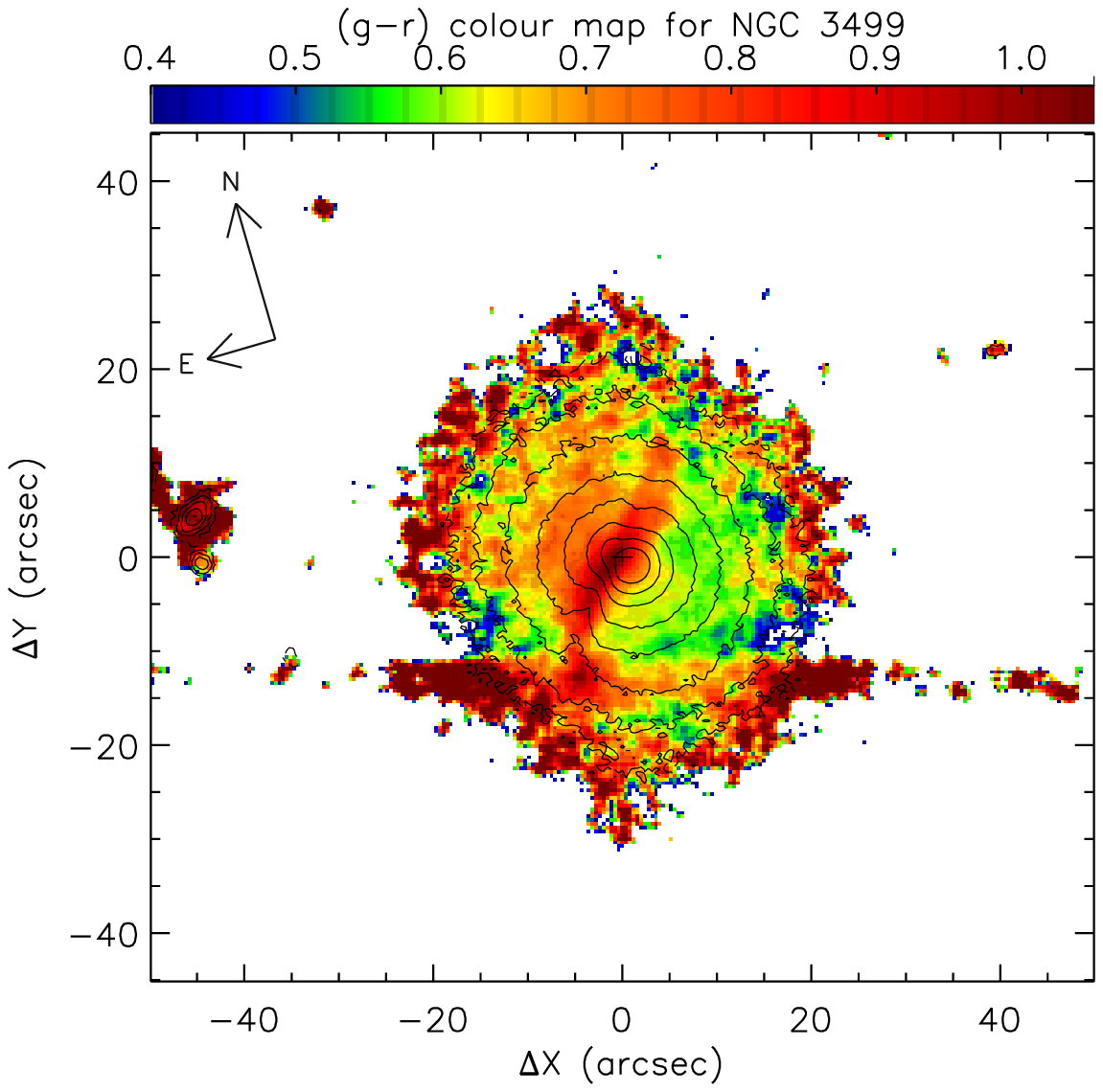} &
 \includegraphics[width=8cm]{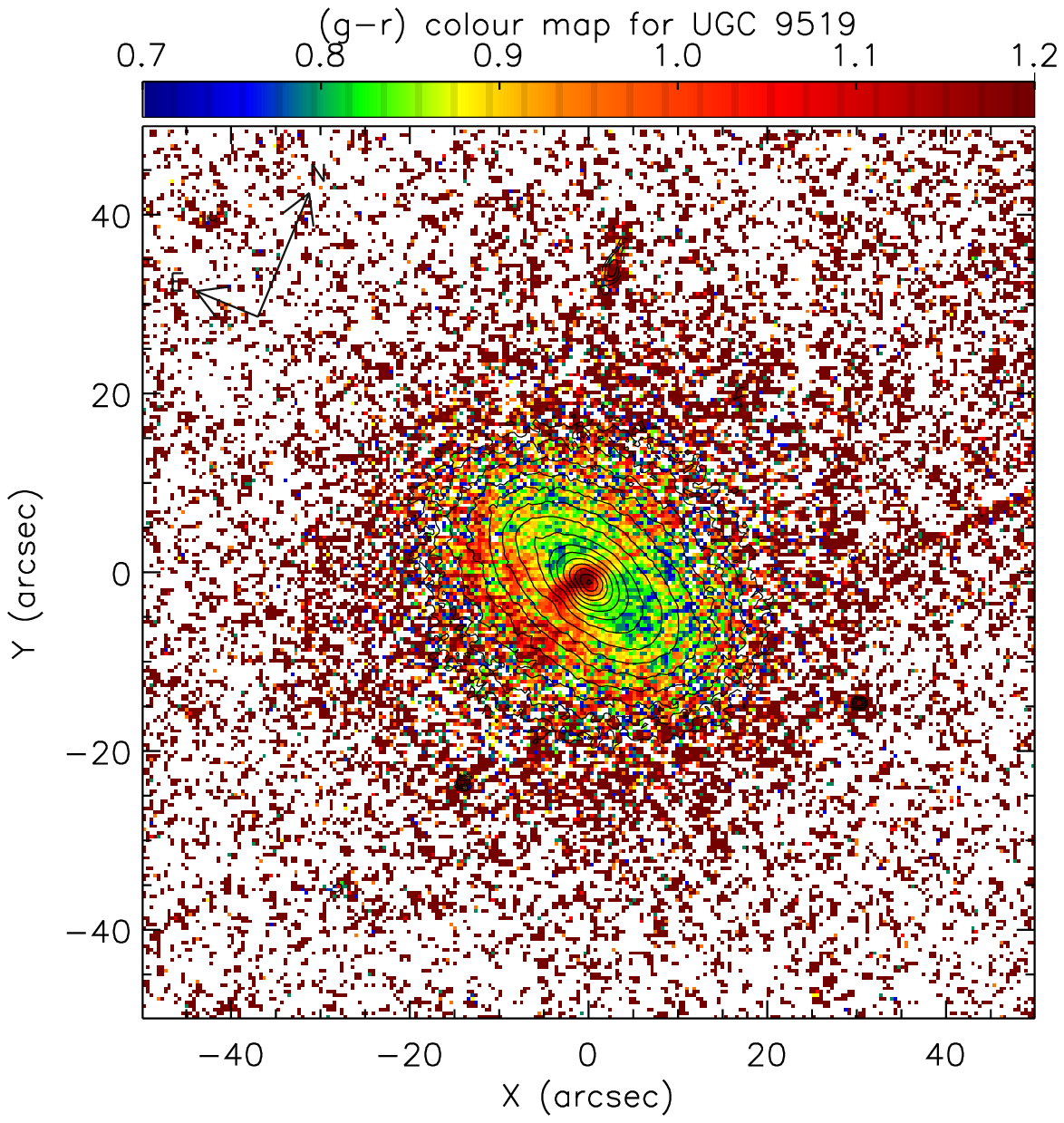} \\
\end{tabular}
\caption{The color maps for NGC~3499 (left) and UGC~9519 (right) derived from the SDSS/DR9 data.}
\label{colmaps}
\end{figure*}

Some examples of the velocity fields for the stellar and gaseous components in the galaxies with
the inner polar disks are shown in Fig.~\ref{velfields}. As a confirmation, or visual signatures
of the inner polar disk presence, especially in the cases when the gaseous disks are seen edge-on, 
I use also the maps of the emission-line intensities -- as the illustrations, the [OIII]$\lambda$5007 
maps for NGC~5507 and NGC~4233 are presented in Fig.~\ref{emismaps}, -- and the color maps constructed 
by using the SDSS archive (DR9) data -- the examples for UGC~9519 and NGC~3499
are given in Fig.~\ref{colmaps}. I note here that the polar inner disks in NGC~4233 and UGC~9519 were
reported earlier, by \citet{polars0} in the former and by \citet{isomnras} in the latter galaxy.
As for UGC~9519, the ATLAS-3D team reported also the external source of the misaligned cold neutral gas for
this galaxy \citep{atlas3d_10}. 

Below some details of the gas distributions and motions in individual galaxies discussed here are specified.

\begin{figure}
\begin{tabular}{c c}
 \includegraphics[width=4cm]{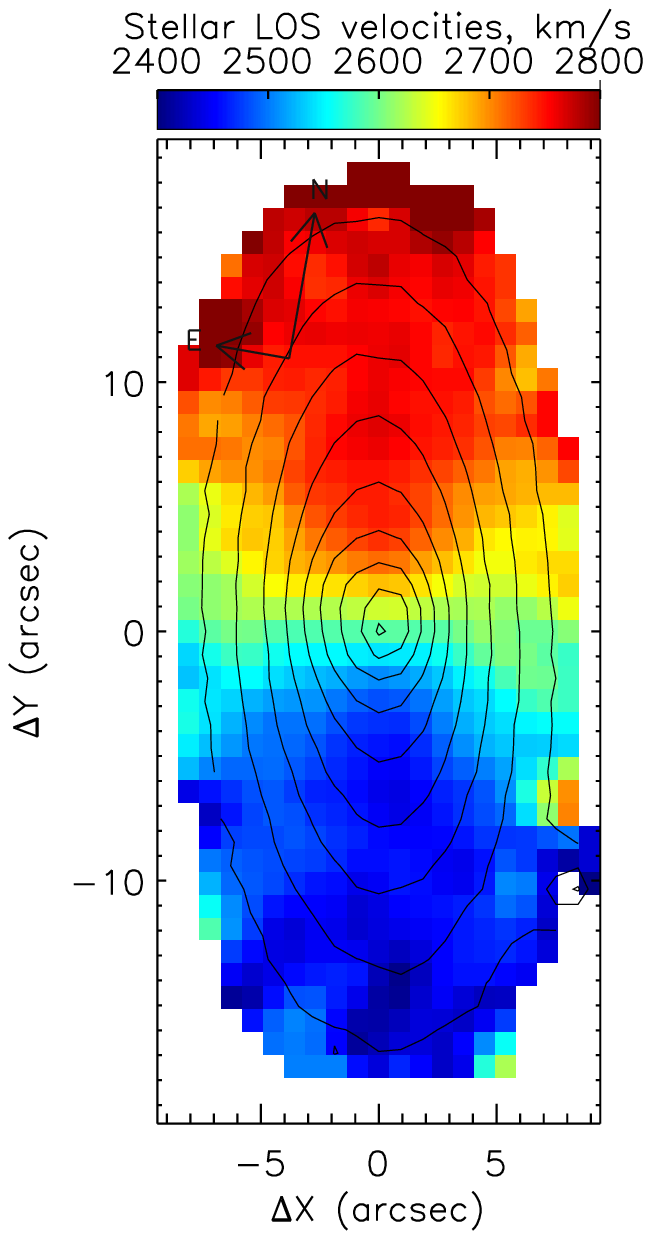} &
 \includegraphics[width=4cm]{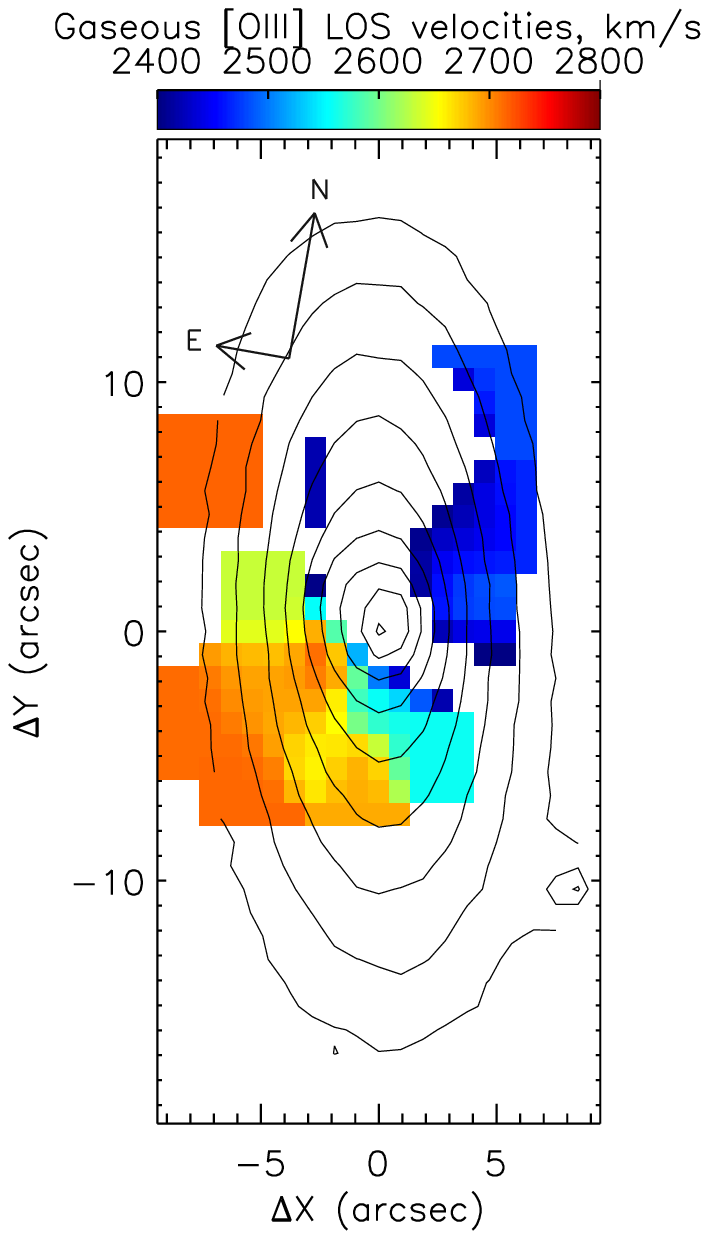} \\
 \includegraphics[width=4cm]{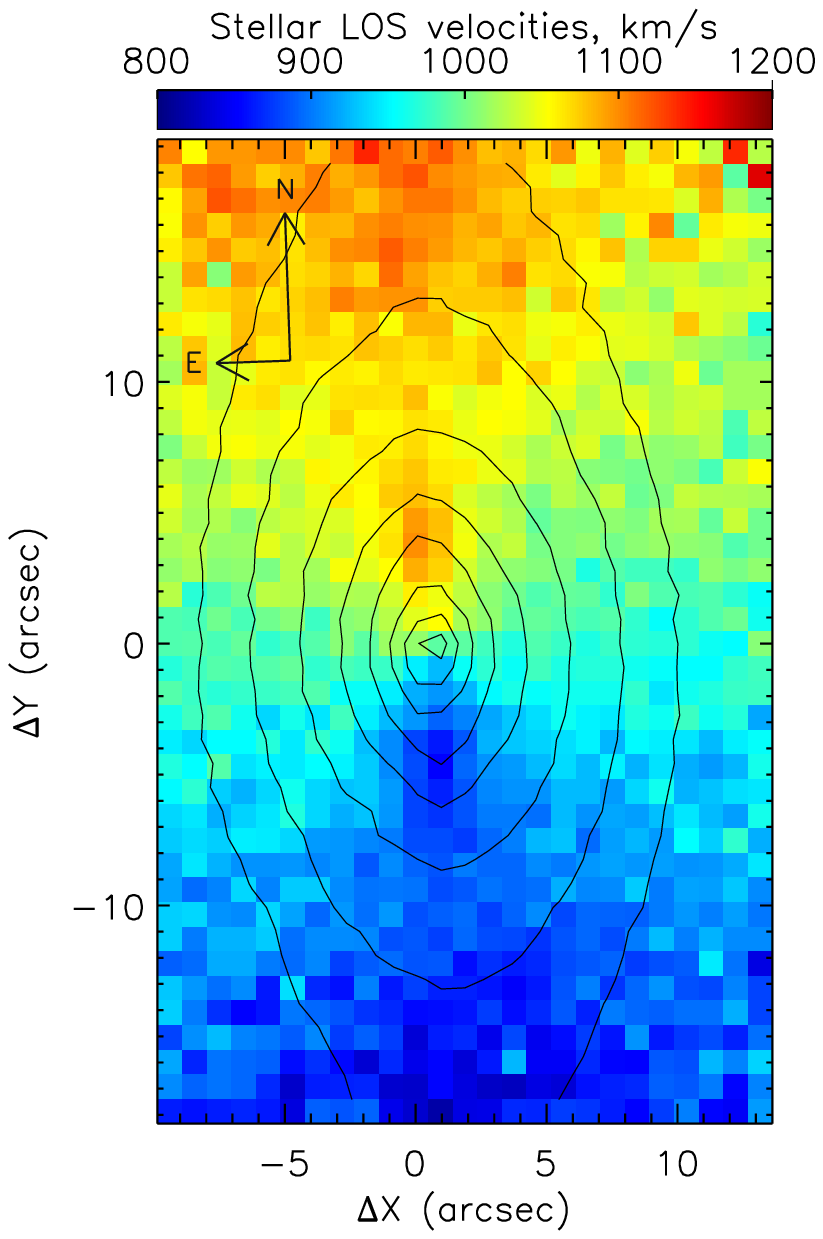} &
 \includegraphics[width=4cm]{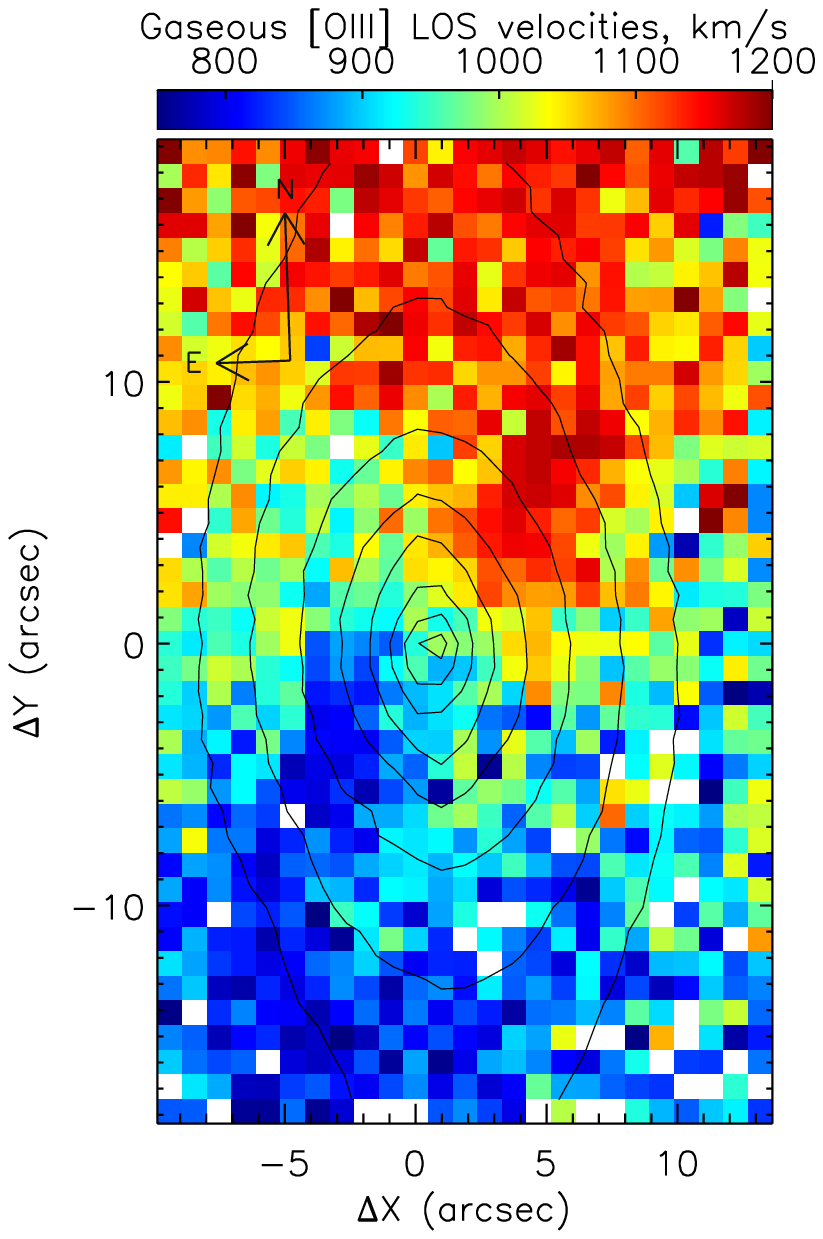} \\
\end{tabular}
\caption{The stellar and gaseous LOS velocity fields of two edge-on galaxies: NGC 1121 (upper plots)
and NGC 4026 (bottom plots). In the central parts the gaseous disks are inclined to the main galactic 
planes by some 50 deg.}
\label{edgeon}
\end{figure}

\noindent
{\bf NGC~1121.} The circumnuclear gaseous subsystem of this galaxy seen in the [OIII]$\lambda$5007
emission line is too compact to be analyzed with the DETKA software. However it is the case when
the stellar disk of the galaxy is seen edge-on, and the centrally concentrated gas rotates
evidently off the main galactic plane (Fig.~\ref{edgeon}, upper plots).

\noindent
{\bf NGC~2962.} This bona fide S0 galaxy is however very gas-rich: \citet{grossi_hi} report some $10^9$ solar
masses of neutral hydrogen in NGC~2962 distributed regularly in an extended disk co-spatial with
the stellar one. But in the center (Fig.~\ref{velfields}, upper left) we see several nested gaseous
subsystems with decoupled rotation: though all kinematical major axes are roughly perpendicular to the
isophote major axis (by some 80 deg), the very inner ionized gas, at $R<4^{\prime \prime}$, counterrotates with 
respect to the more outer gas. At $R=5^{\prime \prime}-18^{\prime \prime}$ the gas rotation plane is strongly
inclined, but not strictly polar, and undergoes smooth warp along the radius. 

\noindent
{\bf NGC~3499.} It is isolated \citep{2mig}, small, close to face-on S0/a galaxy. Its HI content is rather
modest but detectable: $7 \times 10^7$ solar masses according to the Green Bank Telescope integrated data
\citep{wei}. These authors -- \citet{wei} -- describe the neutral hydrogen distribution in NGC~3499 as
a regularly rotating extended disk, with the rotation velocity of 119~$\sin i$ $\km$. We see orthogonality
of the kinematical major axes of the stellar and ionized-gas components in the center of the galaxy 
($PA_*=38^{\circ}\pm 4^{\circ}$ vs $PA_{gas}=133^{\circ}\pm 3^{\circ}$, Fig.~\ref{velfields}, upper right), 
and also a dust lane aligned in $PA=132^{\circ}$, just as the 
kinematical major axis of the ionized gas (Fig.~\ref{colmaps}) but projected off the nucleus; so the central 
gaseous disk is not strictly edge-on.

\noindent
{\bf NGC~3648.} The galaxy lacks HI, according to \citet{atlas3d_13}. So we can suggest a compact circumnuclear
ionized-gas disk, warped and becoming nearly polar at $R>5^{\prime \prime}$ (Table~3).

\noindent
{\bf NGC~4026.} The galaxy is strictly edge-on. In the central part the gaseous component looks like
a disk seen edge-on, inclined by some 50 deg to the main galactic plane. Farther from the center, the gaseous
disk warps and lies into the galactic plane. The HI-map presented by \citet{atlas3d_13} demonstrates
a long gas-rich filament entering into the disk of the galaxy just near the center at the right angle.
Perhaps, we see here the nearly polar gas accretion through the cosmological filament and strong plane
precession over the outer parts of the extended gaseous disk.

\noindent
{\bf NGC~4690.} The galaxy lacks detections both of neutral \citep{bettoni03} and molecular gas \citep{atlas3d_4}.
In our maps (Fig.~\ref{velfields}, bottom right) we see a sort of ionized-gas ring starting at a radius of
$R\approx 2^{\prime \prime}$, extending toward $R\approx 8^{\prime \prime}$, with a possible warp at the edge,
with the PA$_{kin}$(gas) changing by some 55 deg over the radius range of 3 arcsec. The outer, 
$R=7^{\prime \prime}-8^{\prime \prime}$, orientations of the disk lines of nodes, PA$_{kin}$(stars) and PA$_{kin}$(gas), 
are strictly orthogonal.

\noindent
{\bf NGC~5507.} It is also the galaxy where we can suggest a source of the polar gas accretion: it 
has a luminous late-type neighbor, NGC~5506, in 25 kpc from it. The circumnuclear ionized gaseous disk of NGC~5507 
seen in [OIII]$\lambda$5007 is edge-on and roughly west-east elongated (Fig.~\ref{emismaps}), just as NGC~5506 as a
whole. However, the kinematical major axis of this disk is oriented in $PA=138^{\circ}$ so betraying radial gas 
motions, most probably -- inflow: the superposition of the circular rotation having the disk line of nodes aligned in 
$PA\approx 90^{\circ}$ with pure radial motions of comparable velocity amplitude would result in 
kinematical major axis turn by $45^{\circ}$.

Besides these seven S0 galaxies with the polar inner disks found so far in the ATLAS-3D sample, 
we can refer also to the inner polar disks in UGC~9519 implied by the polar dust lane (Fig.~\ref{colmaps}, right) and gas 
decoupled kinematics, see \citet{isomnras}, and in the strongly emissive NGC~4684 where the stellar disk is seen edge-on, 
and the circumnuclear gas is aligned at the direct angle with respect to the main galactic body \citep{atlas3d_10}.
When we expand the present sample of 143 S0s toward the full sample of the ATLAS-3D survey including
early observations in the frame of the SAURON project \citep{sauron}, we obtain 200 lenticular galaxies in the 
local volume limited by $D<42$~Mpc; according to \citet{atlas3d_1}, this sample is complete over the galaxy 
luminosities exceeding $M_K<-21.5$. By adding to the 9 inner polar disks mentioned above also the galaxies listed 
by \citet{moisrev}, we claim the detection of 21 inner polar disks among the full sample of
200 nearby lenticulars. So, the frequency of strongly inclined inner gaseous disks in the nearby S0s is about
10\%. We must note that it is much higher than the incidence of large-scale polar rings, which is
$<1$\% \citep{resh11,pringcat}.

\section{Do the inner polar disk hosts differ from others?} 

\begin{figure}
\begin{tabular}{c c}
 \includegraphics[width=4cm]{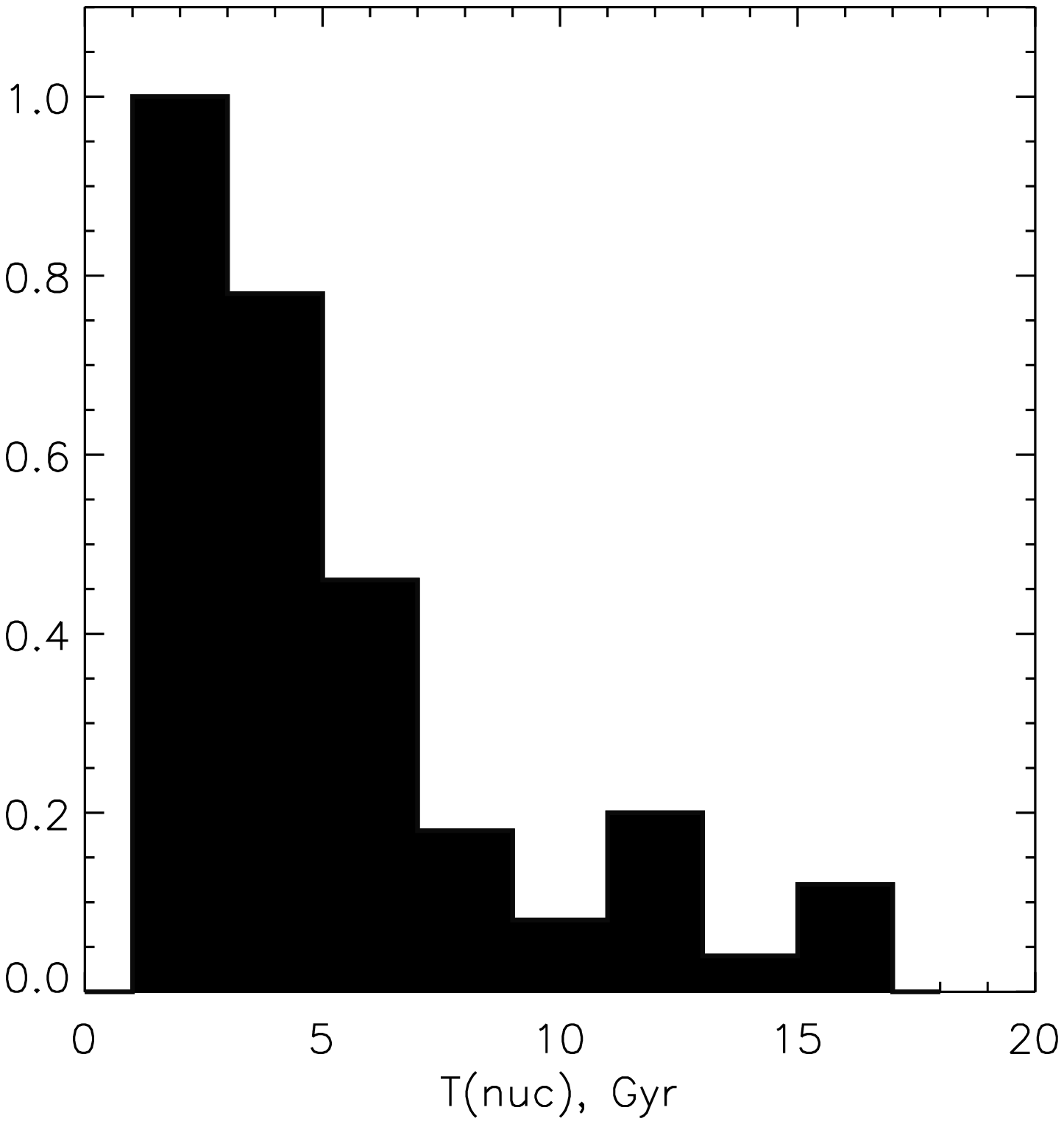} &
 \includegraphics[width=4cm]{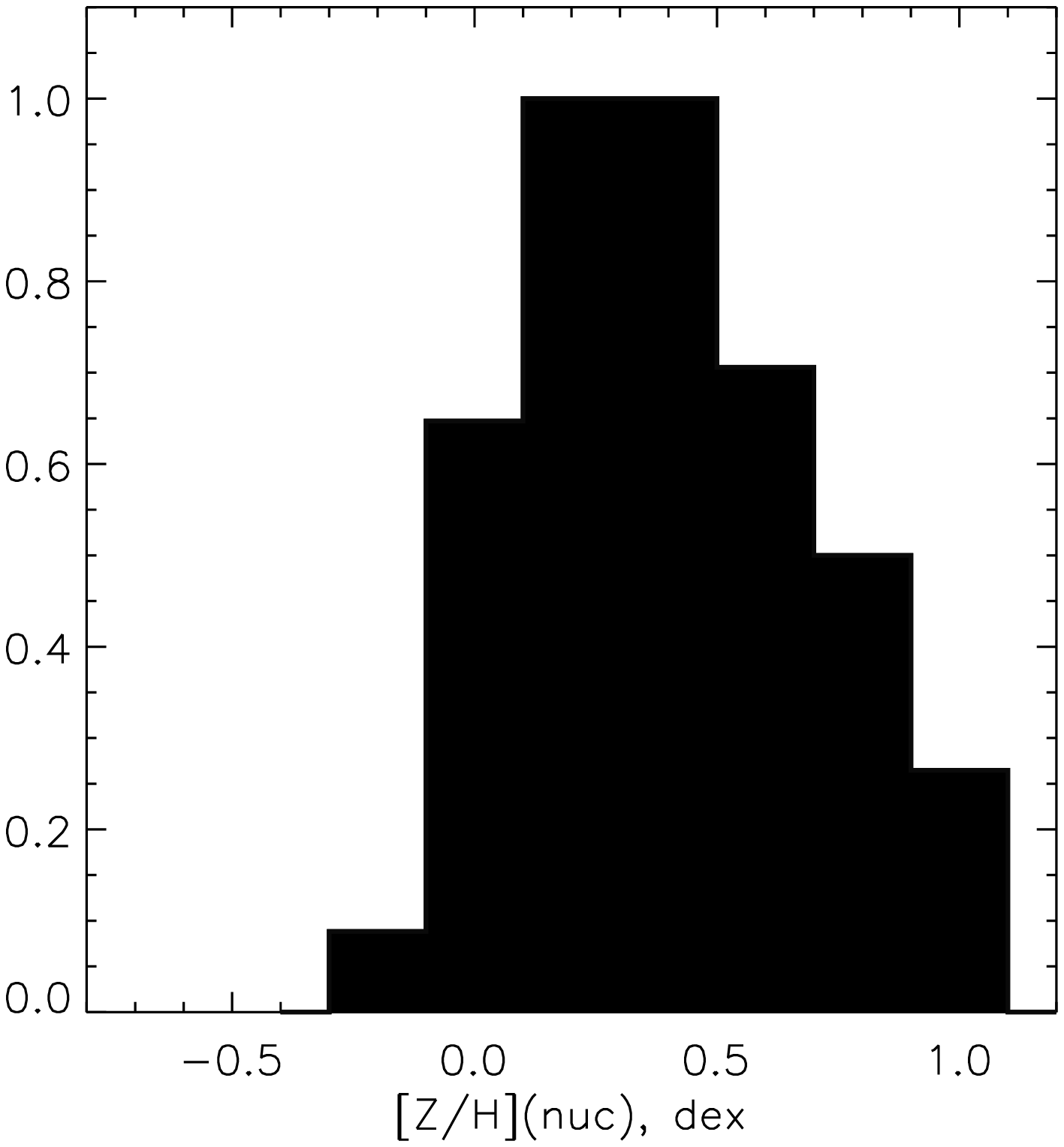} \\
 \includegraphics[width=4cm]{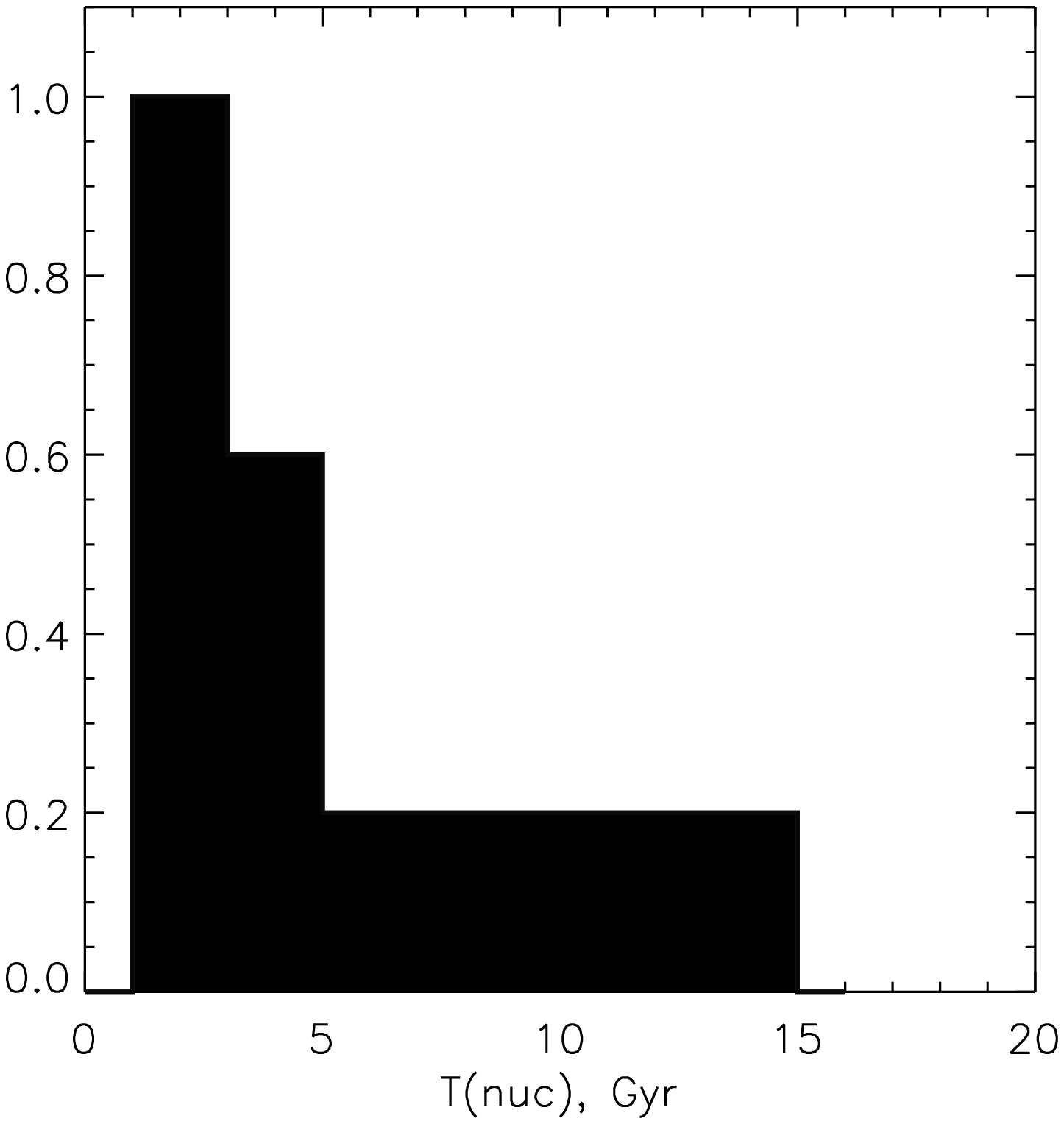} &
 \includegraphics[width=4cm]{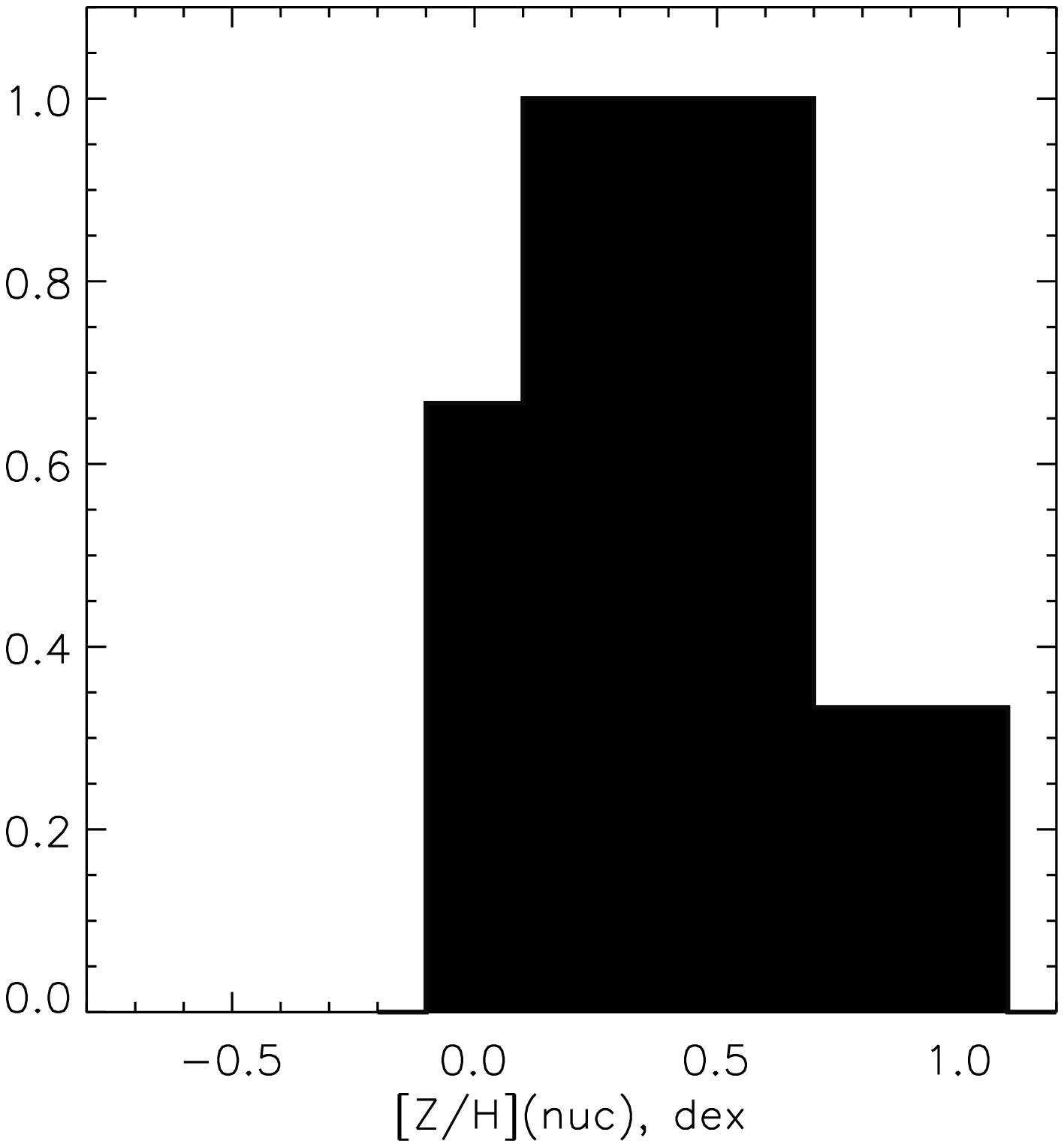} \\
\end{tabular}
\caption{The distributions of the SSP-equivalent characteristics of the nuclear stellar populations
in the large sample of the ATLAS-3D S0s (top row) and in the inner polar disk hosts (bottom
row). The age distributions are to the left and the metallicity distributions -- to the right.}
\label{age_z_histo}
\end{figure}

\begin{figure}
\begin{tabular}{c c}
\includegraphics[width=4cm]{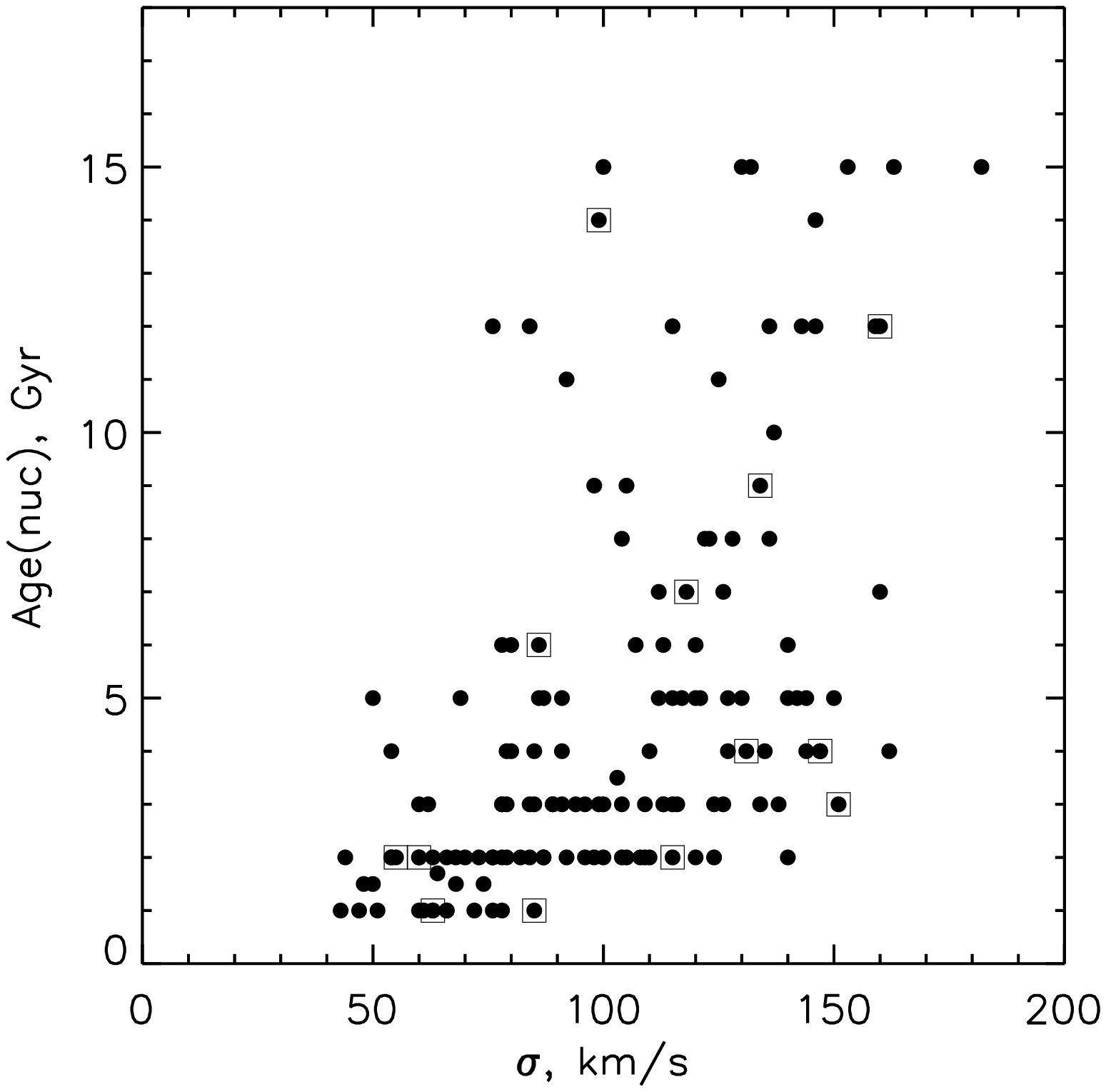} &
\includegraphics[width=4cm]{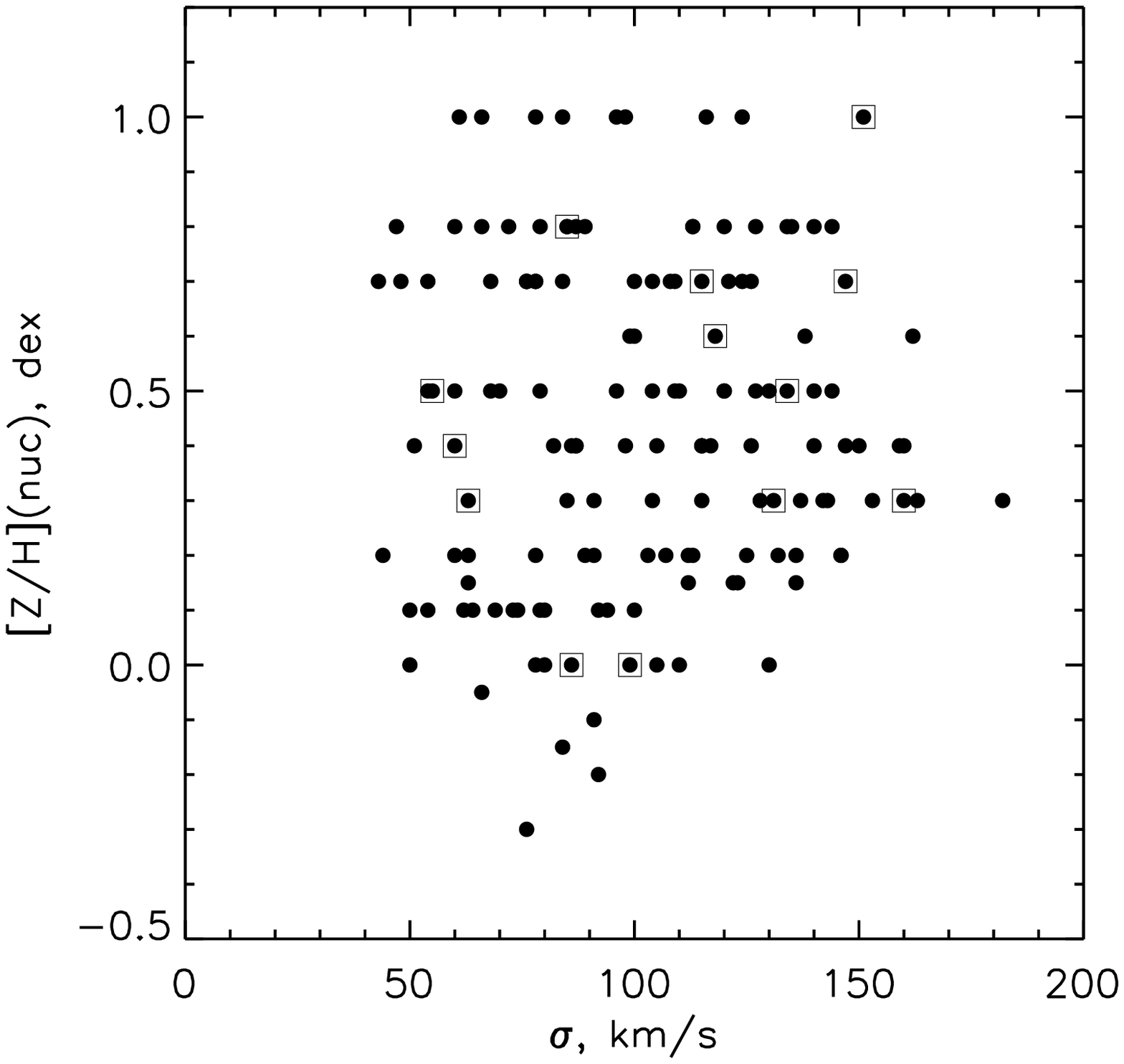} \\
\end{tabular}
\caption{The correlations of the nuclear stellar population SSP-equivalent ages (left) and metallicities
(right) with the central stellar velocity dispersion. The points corresponding to the inner polar disk
hosts are circled in the squares.}
\label{agez_vs_sig}
\end{figure}

Though many lenticular galaxies contain cold gas \citep{welchsage03,welchsage06,welchsage10,atlas3d_10,atlas3d_27},
only for gaseous subsystems with decoupled kinematics there exist commonly accepted view that they have been 
accreted from outside. Among the gaseous subsystems with decoupled kinematics, the polar disks are thought
to appear in a host galaxy by capturing the gas with orthogonal orbital momentum from somewhere outside -- 
see e.g. \citet{polarform,brook08}. Moreover, polar gas, though accreted from outside, is considered to be 
dynamically stable and to inhabit long-lived orbits in S0 galaxies \citep{polarform,mapelli}.
While the outer gas captured into the main galactic plane of a host galaxy must sink by spiralling into the center
on a timescale of 1~Gyr \citep{walker96}, and is expected to feed the central star formation burst with
subsequent rejuvenation of the nuclear stellar population, the polar gas may remain off the nucleus and
avoid affecting nuclear stellar population properties.

Having the complete sample of nearby S0s, among which 10\%\ of the galaxies possess inner polar disks, I can
compare the properties of the nuclear stellar populations in the inner polar disk hosts with the mean
characteristics of the whole population of nearby S0s. Figure~\ref{age_z_histo} shows the distributions
of the SSP-equivalent ages and metallicities of the nuclear stellar populations for the whole sample of
143 S0s and separately for the hosts of the inner polar disks -- 7 new ones, 6 known previously \citep{moisrev}.
The total nuclear age distribution is very specific, with the most galaxies concentrating around the nucleus 
age of $T=2-3$~Gyr, while the minority have old nuclei. The shape of the nucleus-age distribution for the galaxies
with the inner polar disks repeats the general distribution perfectly (the Kolmogorov-Smirnov criterium is
equal to 0.41): we cannot say that the nuclei in
the hosts of the inner polar disks are older or younger on average. The most nuclei, 8 of 13, in the S0s
with the inner polar disks are rather young, of 1--4~Gyr old. Their range of metallicity, [Z/H]$\ge 0.0$,
fits perfectly the total distribution of the nuclear stellar population metallicities in the nearby S0s. 
We can conclude that the evolution of the central parts of the inner polar disk hosts is quite the same as
that of the whole population of S0s: mainly with a secondary nuclear star formation burst at redshift
of $z < 0.5$ ($< 5$~Gyr ago).

Figure~\ref{agez_vs_sig} presents relations between the age or metallicity of the nuclear stellar 
populations and the stellar velocity dispersion averaged over apertures of $6.5^{\prime \prime}$
so characterizing the bulge masses. The hosts of the inner polar disks are encircled by open squares,
to separate out their positions among the total distributions. In general, early-type galaxies reveal 
tight correlations between these characteristics: [Z/H] correlates with $\sigma _*$ for the cluster early-type
galaxies \citep{trager00_2}, while the ages correlate with $\sigma _*$ for samples of field galaxies
starting from group environments \citep{trager00_2,caldwell,howell}. In our data we see that the hosts
of the inner polar disks spreading over the full range of $\sigma _*$ participate in the total relations
which resemble in general the relations for field early-type galaxies.
However, the relation of age vs $\sigma _*$ in Fig.~\ref{agez_vs_sig}
is unexpectedly curious: only young nuclei, $T<7$~Gyr, reveal the tight correlation of their ages with the masses 
of the bulges. Among the old nuclei, no connection between the ages of the nuclear stellar populations and the
masses of the bulges can be found. The nuclei of the hosts of the inner polar disks share this behavior:
10 younger nuclei distributed between $\sigma _* =$ 50 and 150 $\km$ demonstrate good correlation of
their ages vs $\sigma _*$, while among the hosts of 3 older nuclei the oldest one is the less massive, it has 
only $\sigma _*=100\,\km$.

\section{Conclusions}

I have inspected the central stellar population properties and also the gaseous and stellar kinematics 
of the volume-limited sample of nearby S0 galaxies observed with the IFU SAURON in the frame of the 
ATLAS-3D project. I have found seven (7) new cases of nearly polar rotation of the circumnuclear warm-gas 
disks with respect to the stars. Together with these new findings, I report
the presence of 21 inner polar disks among the complete sample of 200 nearby S0s. It means that
the frequency of the inner polar gas rotation is about 10\%\ for the early-type disk galaxies in the local
Universe that is much higher than the incidence of large-scale polar rings which is $<1$\%  \citep{pringcat}. 
Perhaps, this difference reflects the natural difference between frequencies of minor and major mergers. To my surprise,
the properties of the nuclear stellar populations of the inner polar disk hosts are strictly the same
as those of the whole sample of nearby S0s. In particular, the age distributions are quite similar, 
with the most galaxies concentrated around the value of the nuclear stellar age of 1--4~Gyr. 
It means that despite proposed stability of polar orbits, the gas reached the very centers and
provoked recent star formation bursts in the nuclei of the inner polar disk hosts, as it took place
in the majority of S0 galaxies. It remains to be understood if the S0 galaxies suffered multiple
gas accretion events, with only a single one of them from a polar orbit.

\acknowledgments

The present study makes use of data obtained from the Isaac Newton Group Archive which is maintained 
as a part of the CASU Astronomical Data Centre at the Institute of Astronomy, Cambridge, UK.
This research is also partly based on SDSS data.
Funding for the Sloan Digital Sky Survey (SDSS) and SDSS-II has been
provided by the Alfred P. Sloan Foundation, the Participating Institutions,
the National Science Foundation, the U.S. Department of Energy, the National
Aeronautics and Space Administration, the Japanese Monbukagakusho,
and the Max Planck Society, and the Higher Education Funding Council
for England. The SDSS Web site is http://www.sdss3.org/.
The whole work was supported by the Russian Science Foundation (the project number 14-22-00041).

\end{document}